\documentclass{jpp}

\usepackage{graphicx}
\usepackage{natbib}
\usepackage{amsmath}
\usepackage{amssymb}
\usepackage{mathtools}
\usepackage{lscape}
\usepackage{color}  
\usepackage{rotating}
\usepackage{caption}
\usepackage{ulem}
\usepackage{psfrag}
\usepackage{pstool}
\usepackage[colorlinks,urlcolor=blue,citecolor=blue,linkcolor=blue]{hyperref}
\usepackage{multirow}

\ifCUPmtlplainloaded \else
  \checkfont{eurm10}
  \iffontfound
    \IfFileExists{upmath.sty}
      {\typeout{^^JFound AMS Euler Roman fonts on the system,
                   using the 'upmath' package.^^J}%
       \usepackage{upmath}}
      {\typeout{^^JFound AMS Euler Roman fonts on the system, but you
                   dont seem to have the}%
       \typeout{'upmath' package installed. JPP.cls can take advantage
                 of these fonts, if you use 'upmath' package.^^J}%
      }
  \else
  \fi
\fi

\ifCUPmtlplainloaded \else
  \checkfont{msam10}
  \iffontfound
    \IfFileExists{amssymb.sty}
      {\typeout{^^JFound AMS Symbol fonts on the system, using the
                'amssymb' package.^^J}%
       \usepackage{amssymb}%
       \let\le=\leqslant  \let\leq=\leqslant
       \let\ge=\geqslant  \let\geq=\geqslant
      }{}
  \fi
\fi

\ifCUPmtlplainloaded \else
  \IfFileExists{amsbsy.sty}
    {\typeout{^^JFound the 'amsbsy' package on the system, using it.^^J}%
     \usepackage{amsbsy}}
    {}
\fi

\providecommand\sgn{\text{sgn}}

\newsavebox{\astrutbox}
\sbox{\astrutbox}{\rule[-5pt]{0pt}{20pt}}

\def\<{\langle}
\def\>{\rangle}

\def\bi{\bf}
\def\rmd{d}
\def\rmi{i}

\providecommand{\av}[1]{\langle{#1}\rangle}
\providecommand{\aV}[1]{\left\langle{#1}\right\rangle}

\def\be{\begin{equation}}
\def\ee{\end{equation}}
\def\bea{\begin{eqnarray}}
\def\eea{\end{eqnarray}}
\def\nn{\nonumber}
\newcommand{\ms}{\noalign{\vspace{3pt plus2pt minus1pt}}}

\newfont{\myfont}{cmmib10}

\title[Wave dispersion in pulsar plasma 1]{Wave dispersion in pulsar plasma:\\ 1. Plasma rest frame}

\author[M. Z. Rafat, D. B. Melrose and A. Mastrano]%
{M.\ns Z. \ns R\ls A\ls F\ls A\ls T$^{1}$ \ns
D.\ns B.\ns M\ls E\ls L\ls R\ls O\ls S\ls E$^1$%
  \thanks{Email address for correspondence: donald.melrose@sydney.edu.au},\ns \and
A.\ns  M\ls A\ls S\ls \ls T\ls R\ls A\ls N\ls O$^{1}$}
\affiliation{$^1$SIfA, School of Physics, The University of Sydney, NSW 2006, Australia
}

\pubyear{2017}
\volume{?}
\pagerange{?}
\date{?; revised ?; accepted ?. - To be entered by editorial office}
\begin{document}

\maketitle

\begin{abstract}
Wave dispersion in a pulsar plasma (a 1D, strongly magnetized,  pair plasma streaming highly relativistically with a large spread in Lorentz factors in its rest frame) is discussed, motivated by interest in beam-driven wave turbulence and the pulsar radio emission mechanism. In the rest frame of the pulsar plasma there are three wave modes in the low-frequency, non-gyrotropic approximation. For parallel propagation (wave angle $\theta=0$) these are referred to as the X, A and L modes, with the X and A modes having dispersion relation $ |z| = z_{\rm A}\approx1-1/2\beta_{\rm A}^2$, where $ z = \omega/k_\parallel c$ is the phase speed and $\beta_{\rm A}c$  is the Alfv\'en speed. The L~mode dispersion relation is determined by a relativistic plasma dispersion function, $z^2W(z)$, which is negative for $|z| < z_0$ and has a sharp maximum at $|z| = z_{\rm m}$, with $1-z_{\rm m}<1-z_0\ll1$. We give numerical estimates for the maximum of $z^2W(z)$ and for $z_{\rm m}$ and $z_0$ for a 1D J\"uttner distribution. The L and A~modes reconnect, for $z_{\rm A}>z_0$,  to form the O and Alfv\'en modes for oblique propagation ($\theta\neq0$). For $z_{\rm A}<z_0$ the Alfv\'en and O~mode curves reconnect forming a new mode that exists only for $\tan^2\theta \gtrsim z_0^2-z_{\rm A}^2$. 

The L~mode is the nearest counterpart to Langmuir waves in a nonrelativistic plasma, but we argue that there are no ``Langmuir-like'' waves in pulsar plasma, identifying three features of the L~mode (dispersion relation, ratio of electric to total energy and group speed) that are not Langmuir-like. A beam-driven instability requires a beam speed equal to the phase speed of the wave. This resonance condition can be satisfied for the O~mode, but only for an implausibly energetic beam and only for a tiny range of angles for the O~mode around $\theta\approx0$. The resonance is also possible for the Alfv\'en mode but only near a turnover frequency that has no counterpart for Alfv\'en waves in a nonrelativistic plasma.
\end{abstract}

\begin{PACS}
\end{PACS}

\section{Introduction}

The mechanism by which pulsar radio emission is generated remains controversial. Several different mechanisms continue to attract supporters and critics, including coherent curvature emission (CCE), relativistic plasma emission (RPE), anomalous Doppler emission (ADE), linear acceleration emission (LAE) and free-electron maser emission (FEM). Two of these (RPE and ADE) depend intrinsically on the wave dispersion in the ``pulsar plasma'' which we define to be a one-dimensional (1D), electron-positron plasma, streaming outward at a relativistic velocity $\beta_{\rm s}c$ with streaming Lorentz factor $\gamma_{\rm s}=(1-\beta_{\rm s}^2)^{-1/2}\gg1$, and with an intrinsically relativistic spread in Lorentz factors, $\langle\gamma\rangle\gg1$, where $ \av{\cdots} $ denote an average, in the rest frame of the streaming plasma. Wave dispersion in such a pulsar plasma has major differences from wave dispersion in a conventional plasma, due to the extreme anisotropy, in the form of 1D distributions, the absence of ions, the highly relativistic energies of the bulk of the particles, and the superstrong magnetic field. For example, all waves in a pulsar plasma have phase speeds very close to or above the speed of light, and the longitudinal waves exist only for propagation parallel to the magnetic field.

Renewed interest in RPE has been stimulated by a recent argument in favor of a ``beam-driven'' form of RPE:  \citet{EH16} argued that RPE is the only one of the suggested mechanisms that can plausibly account for nanoshots from the Crab pulsar. Specifically, the form of RPE invoked by \citet{EH16} involves beam-driven Langmuir-like waves, which are assumed to build up to a very high level in localized regions in the pulsar magnetosphere through a mechanism suggested by \citet{W97,W98}. This argument has potentially wider implications: it is intrinsically unlikely that the emission mechanism in nanoshots is unrelated to other pulsar radio emission and hence, if the case for RPE operating in nanoshots is accepted, this would provide a strong argument for RPE being the generic mechanism for pulsar radio emission. In most forms of RPE \citep{SC73,SC75,H76b,H76a,HR76,HR78,BB77,LMS79,LP82,APS83,ELM83,L92,A93,A95,W94}, including the form proposed for nanoshots, it is assumed that Langmuir-like waves exist in the pulsar plasma, with properties similar to Langmuir waves in a nonrelativistic plasma, but there are no such waves in a pulsar plasma. Any growing wave must be in a specific wave mode of the pulsar plasma.

In this paper, and in two accompanying papers (referred to as Papers~2 and~3), we discuss wave dispersion and beam-driven instabilities in pulsar plasma with the objective of providing a systematic description of the underlying plasma theory needed in a critical discussion of the beam-driven instabilities invoked in RPE and CCE. Our main purpose in this paper is to describe wave dispersion in a plasma with the properties that we postulate here for a pulsar plasma in its rest frame. In particular we emphasize the importance of the relativistic spread $\langle\gamma\rangle\gg1$.  In all three papers we assume that every particle distribution is a one-dimensional J\"uttner distribution, which is of the form $g(u)\propto\exp(-\rho\gamma)$, with $\rho$ the inverse temperature in units of the rest energy of the electron, and with $u=\gamma\beta$ the 4-speed. Numerical models suggest that the pair cascade results in broad particle distributions  \citep{HA01,AE02,ML10,TA13}, which have been described as J\"uttner-like  \citep[e.g.,][]{AE02}.

A J\"uttner distribution should be regarded as the default choice for the distribution function for the particles in a pulsar plasma and in other relativistic astrophysical pair plasmas. A J\"uttner distribution \citep{J11,Synge57,WH75} is the relativistic generalization of a thermal (or Maxwellian) distribution. We suggest that a pulsar plasma should be regarded as analogous to most nonrelativistic astrophysical plasmas in the sense that there is a ``background'' distribution that is thermal (Maxwellian or J\"uttner) with suprathermal tails and other nonthermal features regarded as complementary distributions or as modifications to this ``background'' distribution. The wave dispersion is assumed to be determined primarily by the ``background'' with instabilities attributed to nonthermal features.

Indirect evidence for J\"uttner distribution follows from numerical calculations, particularly particle-in-cell calculations, from which the form of the distribution function can be inferred. As already noted, the results of such calculations for pair creation in a pulsar plasma suggest a 1D J\"uttner distribution. Another astrophysical problem concerns the propagation of a shock into an electron-positron plasma, and again the resulting distribution of the post-shock pair plasma is consistent with a J\"uttner distribution \citep[e.g.,][]{Gallant_etal92,SS09,Iwamoto17}.  While such calculations do not provide compelling evidence for a J\"uttner distribution, they do support the suggestion that the default choice should be a J\"uttner distribution. One needs a specific reason for choosing any other distribution for a relativistic pair plasma in an astrophysical context. The implications of this default assumption are more important for streaming distributions (Paper~2) than for the non-streaming distribution considered in the present paper, where the main effect is the large intrinsic spread in Lorentz factors, $\langle\gamma\rangle\approx1/\rho\gg1$ for the highly relativistic case, $\rho\ll1$, of interest here. \cite{AE02} suggested a J\"uttner distribution with $ \rho \approx 1 $ and $ \gamma_{\rm s} = 10^2\text{--}10^3 $ as being relevant to pulsars. Such numerical models apply in local regions where pair cascades occurs, and it is plausible that many such local regions contribute to the distribution function of the outwardly streaming pulsar plasma of interest here. A distribution consisting of many such local distributions has a wide spread in Lorentz factor, and we suggest that this may be modelled by a J\"uttner distribution with a smaller value of $ \rho $. We discuss both the case $ \rho = 1 $, and also the case $ \rho \ll 1 $. We find that the wave dispersion for $ \rho = 1 $, is more closely analogous to the highly relativistic case $ \rho \ll 1 $ than to the nonrelativistic case $\rho\gg1$.

There are three wave modes in a pulsar plasma, labeled here as X, L and A for parallel propagation and as X, O and Alfv\'en for oblique propagation. It is convenient to choose the phase speed $z=\omega/k_\parallel c$ as the independent variable, and to introduce two plasma parameters: the plasma frequency, $\omega_{\rm p}$, defined without including any Lorentz factors, and $\beta_{\rm A}$ which is such that $\beta_{\rm A}c$ is the Alfv\'en velocity, as conventionally defined, with $\beta_{\rm A}\gg1$ in a pulsar magnetosphere. For parallel propagation, the X~and A~modes are degenerate with dispersion relation $z=z_{\rm A}$, with $z_{\rm A}\approx\beta_{\rm A}/(1+\beta_{\rm A}^2)^{1/2}$. The L~mode has dispersion relation $\omega=\omega_{\rm L}(z)$, with $\omega_{\rm L}^2(z)=\omega_{\rm p}^2z^2W(z)$, where $z^2W(z)$ is a relativistic plasma dispersion function (RPDF), with $ W(z) $ defined in~\eqref{Wz}.

RPE may be regarded as a pulsar counterpart of conventional plasma emission, for example, in solar radio bursts \citep[e.g.][p. 94]{M86}. Plasma emission involves at least two stages, with the first stage being an instability that generates turbulence in Langmuir waves, and another stage involving partial conversion of energy in this turbulence into escaping radio waves. Two difficulties were recognized in early discussions of RPE. First, the growth rate for various suggested instabilities in the first stage is too slow to be effective, and ways in which this might be overcome were proposed and explored \citep{U87,UU88,AM98,GMG02}. Second, the conversion mechanism into escaping radiation is problematic, and was referred to as a ``bottle-neck'' by \citet{U00}. Here we are concerned with a third difficulty with RPE: the existence of ``Langmuir-like'' waves that  can be generated through a beam instability. This difficulty is obscured in most discussions of RPE through over-simplified or implicit assumptions about the wave dispersion. Examples include the assumption that the plasma is cold or nonrelativistic (in its rest frame), or that the waves have nonrelativistic phase and/or group speeds. We argue that in a relativistic pair plasma, with $\rho\lesssim1$, all waves have relativistic phase speeds, calling into question the possibility of beam-driven ``Langmuir-like'' waves in pulsar plasma. We conclude that, like other proposed emission mechanisms, beam-driven RPE encounters seemingly overwhelming difficulties as the generic pulsar radio emission mechanism.

In \S\ref{sect:dispersion} we present a general theory for wave dispersion in a pulsar plasma, and in \S\ref{sect:RPDF} we summarize the properties of the RPDF. In \S\ref{sect:cold} we describe the wave dispersion in cases where the spread $\langle\gamma\rangle$ is neglected (``cold'' plasma, $\langle\gamma\rangle=1$), where it is nonrelativistic ($\langle\gamma\rangle-1\ll1$), and where relativistic effects are important ($ \av{\gamma} \gtrsim 1 $). In \S\ref{sect:rhogg1} we derive results for the wave dispersion in the plasma rest frame for $\langle\gamma\rangle\gg1$. In \S\ref{sect:Lmode} we discuss the properties of the L~mode in more detail, emphasizing the reasons why the mode should not be regarded as ``Langmuir-like''. We discuss our results and summarize our conclusions in \S\ref{sect:conclusions}.

\section{Wave dispersion in pulsar plasma}
\label{sect:dispersion}

In this section we discuss wave dispersion in a pulsar plasma based on the approach presented by \citet{MGKF99} and \citet{MG99}. We present our detailed calculations and assumptions in Appendix~\ref{app:dispersion}.

\subsection{Wave dispersion in two frames}

Dispersion in any plasma may be described by its dielectric tensor $K_{ij}(\omega,{\bi k})$ and the wave equation written in the form $\Lambda_{ij}(\omega,{\bi k})e_j=0$, with ${\bi e}$ the polarization vector and with wave equation tensor
\be
    \Lambda_{ij}(\omega,{\bi k})
        = \frac{c^2(k_ik_j-|{\bi k}|^2\delta_{ij})}{\omega^2}+K_{ij}(\omega,{\bi k}),
    \label{we1}
\ee
where $ \delta_{ij} $ is the Kronecker delta. The dispersion equation is given by setting the determinant of $\Lambda_{ij}$ to zero. In Appendix~\ref{app:dispersion} we derive the nonzero components of $ \Lambda_{ij}(\omega, {\bi k}) $ as given by \eqref{eq:Lambdaij}, making the low-frequency approximation in Appendix~\ref{app:low_frequency} and the non-gryrotropic approximation in Appendix~\ref{app:non_gyrotropic}.  The dispersion equation then may be written as, $ \Lambda_{ij}(\omega, {\bi k})  \to \Lambda_{ij} $,
\begin{equation}\label{eq:dispersion_equation}
    \det \Lambda_{ij}
        = \Lambda_{22}\left(\Lambda_{11}\Lambda_{33}-\Lambda_{13}^2\right)
        = 0.
\end{equation}
The dispersion relation for any specific wave mode is a specific solution of the dispersion equation~\eqref{eq:dispersion_equation}. Here we derive and discuss the wave properties in the rest frame of the plasma; we discuss the wave properties Lorentz transformed to the pulsar frame in Paper~2. We argue for the use of a J\"uttner distribution to describe a pulsar plasma~\citep{AE02}, which is an even function of $ \beta $ in the plasma rest frame. For a distribution that is an even function of $ \beta $ the nonzero components of $ \Lambda_{ij} $ are given by~\eqref{Lambdaij1}.

For parallel propagation, $ \theta = 0 $, we have $ \Lambda_{13} = 0 $ so that the solutions of~\eqref{eq:dispersion_equation} are $ \Lambda_{22} = 0 $, $ \Lambda_{11} = 0 $ and $ \Lambda_{33} = 0 $ which give the dispersion equation for the parallel X~mode, parallel Alfv\'en or A~mode and the parallel longitudinal or L~mode, respectively, with explicit expressions
\begin{equation}\label{eq:parallel_modes}
    z^2 = z_{\rm A}^2, \quad
    z^2 = z_{\rm A}^2,\quad
    \omega^2 = \omega_{\rm p}^2 z^2 \Re W(z) \equiv \omega_{\rm L}^2(z),
\end{equation}
where $ z_{\rm A} $ is given in~\eqref{Lambdaij1}, $ \Re W(z) $ denotes the real component of $ W(z) $, with $ W(z) $ defined in~\eqref{Wz}. For simplicity in writing we omit $ \Re $ in the definition of $ \omega_{\rm L}^2(z) $. The parallel X and A~modes are identical and may also be expressed as $ \omega^2 = z_{\rm A}^2 c^2 k_\parallel^2 $.

For oblique propagation, $ \theta \neq 0 $, the solution $ \Lambda_{22} = 0 $ gives the oblique X~mode,
\begin{equation}\label{eq:oblique_X_mode}
    z^2 
        = z_{\rm A}^2(1 + \tan^2\theta/b)
    \quad {\rm or} \quad
    \omega^2 
        = z_{\rm A}^2(1 + \tan^2\theta/b) c^2 k_\parallel^2,
\end{equation}
where $ b $ is given in~\eqref{Lambdaij1}. The solution $ \Lambda_{11}\Lambda_{33}-\Lambda_{13}^2 = 0 $ gives the Alfv\'en and the O~modes,
\begin{equation}\label{eq:Alfven_O_modes}
    \omega^2 
        = \frac{(z^2-z_{\rm A}^2)\,\omega_{\rm L}^2(z)}{z^2-z_{\rm A}^2-b\tan^2\theta}
    \quad{\rm or}\quad
    z^2 
        = z_{\rm A}^2+\frac{\omega^2\,b\tan^2\theta}{\omega^2-\omega_{\rm L}^2(z)},
\end{equation}
where $ \omega_{\rm L}(z) $ is given in~\eqref{eq:parallel_modes}. For $ \omega_{\rm L}^2(z) > 0 $, the O~mode is given by~\eqref{eq:Alfven_O_modes} over $ z_{\rm A}^2 + b\tan^2\theta < z^2 \leq \infty $ and the Alfv\'en mode is over $ z_0^2 \leq z^2 < z_{\rm A}^2 $ with $ z_0 $ such that $ \omega_{\rm L}^2(z_0) = 0 $. The case $ \omega_{\rm L}^2(z) < 0 $ is more subtle and is discussed below. 

The polarization vector ${\bi e}$ corresponding to a solution of $ \Lambda_{22} = 0 $ is along the 2-axis, which implies that the X~mode is strictly transverse. Any solution of $ \Lambda_{11}\Lambda_{33} - \Lambda_{13}^2 = 0 $ corresponds to a polarization vector, ${\bf e}$, in the 1-3~plane with
\be
    \frac{e_1}{e_3} = -\frac{\Lambda_{33}}{\Lambda_{13}}=-\frac{\Lambda_{13}}{\Lambda_{11}}.
    \label{LOpol}
\ee
Longitudinal polarization corresponds to ${\bf e}=(\sin\theta,0,\cos\theta)$, and the only strictly longitudinal waves are for parallel propagation, $ \sin\theta = 0 $, satisfying the dispersion equation $ \Lambda_{33} = 0 $, which implies that the L~mode is strictly longitudinal.

There are three modes present for either parallel propagation (parallel X, A and L~modes) or oblique propagation (oblique X, Alfv\'en and O~modes). In the rest frame of the plasma, each mode has a forward-propagating component, $ z > 0 $, and a backward-propagating component, $ z < 0 $. The backward-propagating portion is a mirror image of the forward-propagating one about $ z = 0 $. We restrict our discussion to the forward-propagating solution from which the properties of the backward-propagating part can be readily inferred.

\section{Relativistic plasma dispersion function}
\label{sect:RPDF}

Wave dispersion in a nonrelativistic plasma with Maxwellian distributions of particles may be described in terms of the well-known plasma dispersion function, which has both real and imaginary parts. As usually defined the real part determines the wave dispersion and the imaginary part determines damping of the waves due to resonant absorption. Wave dispersion in a pulsar plasma similarly involves the real and imaginary parts of the RPDF $z^2W(z)$. 

\subsection{1D J\"uttner distribution}

Several different choices have been made for the distribution function of the electrons in a pulsar magnetosphere, including a power-law \citep[][\S17]{KT73}, a relativistically streaming Gaussian distribution \citep[e.g.,][]{ELM83,AM98}, and water-bag and bell distributions \citep{AB86,MG99}. Although the distributions function for the electrons and positrons in the rest frame of the plasma is not known, a relativistic thermal distribution is one approximate form suggested by numerical models for the cascade that generates the pair plasma \citep{HA01,AE02}. We choose a 1D J\"uttner distribution \citep{MG99,MGKF99,AR00} and suggest that this should be the default choice, with a specific reason being required to justify any other choice. (In Paper~2 a streaming distribution is modeled by Lorentz transforming a J\"uttner distribution from its rest frame to the frame in which it is streaming.)

The combined distribution function, for electrons plus positrons, is then
\be
    g(u)=\frac{n\,e^{-\rho\gamma}}{2K_1(\rho)},
    \quad\text{with}\quad
    n = \int_{-\infty}^{\infty} {\rm d}u\,g(u),
    \label{Juttner}
\ee
where $u=\gamma\beta$ and $n$ is the number density. The parameter $\rho=mc^2/T$ is the ratio of the rest energy of the electron to the temperature in energy units, with $\rho=1$ corresponding to $T=0.511\rm\,MeV\approx0.6\times10^{10}\,K$, and $K_1(\rho)$ is the Macdonald function of order 1. One has $\langle\gamma\rangle\approx1/\rho$ for a highly relativistic distribution where $\rho\ll1$.

\subsection{RPDF for a J\"uttner distribution}

The integrand of~\eqref{Wz} defining the RPDF $ W(z) $ is singular at $ \beta = z $ for $ |z| < 1 $. The singularity is treated following the Landau prescription: $\omega \to \omega + i0$ (we assume real $ k_\parallel $). With $z=\omega/k_\parallel c$, this implies $z\to z+i0$ for $k_\parallel>0$ and to $z\to z-i0$ for $k_\parallel<0$. For $\omega>0$, the resonant denominator is then replaced by $ i\pi\, \sgn{(k_\parallel)}\delta(\beta-z) $ where $ \sgn{(k_\parallel)} = k_\parallel/|k_\parallel| $ is the sign of $ k_\parallel $. The RPDF may be expressed as
\be
W(z) = 
\begin{dcases}
	\left\langle\frac{1}{\gamma^3(\beta-z)^2}\right\rangle, & \text{for}\quad |z| > 1,\\
    \frac{1}{n}\bigg[i\pi\left.\sgn{(k_\parallel)}\frac{\rmd g(u)}{\rmd \beta}\right|_{\beta = z} - \left.2\gamma^2 g(u)\right|_{\beta = z}
    &\\\quad\quad\quad\quad\quad
    - \wp\int_{-1}^{1}{\rm d}\beta\, \frac{\left.g(u)\right|_{\beta = z} - g(u)}{(\beta - z)^2}\bigg], & \text{for}\quad |z| \leq 1,
\end{dcases}
\label{Wz2}
\ee
where $ \wp $ denotes a Cauchy Principal Value integral, and $ u  = \gamma\beta = \beta/(1-\beta^2)^{1/2} $.  The expression for $ |z| > 1 $ follows from a partial integration of~\eqref{Wz} and that for $ |z| \leq 1 $ is derived in Appendix~\ref{sect:appRPDF}. Note that the expression for $|z|>1$ can be misleading if applied to $|z|<1$. Specifically, the form for $|z|>1$ is real and positive definite, whereas for $|z|<1$ the RPDF is complex and its real part is negative for $0<|z|\le z_0$, where $z_0$ is identified below. For a J\"uttner distribution the imaginary part of $ W(z) $ follows from~(\ref{Wz2}) with~(\ref{Juttner}) implying
\begin{equation}\label{eq:ImW}
    \Im z^2W(z) = -\sgn{(k_\parallel)}\frac{\pi\rho z^3\gamma_\phi^3 \exp(-\rho\gamma_\phi)}{2K_{1}(\rho)},\quad
    \gamma_\phi = \frac{1}{\sqrt{1 - z^2}}.
\end{equation}

The RPDF $ W(z) $ for the distribution~(\ref{Juttner}) can be expressed in terms of another RPDF,
\be
W(z)=\frac{1}{2K_1(\rho)}\frac{\partial T(z,\rho)}{\partial z},
\qquad
T(z,\rho)=\int_{-1}^{1}{\rm d}\beta\,\frac{e^{-\rho\gamma}}{\beta-z}.
\label{Tzrho}
\ee
The properties of the RPDF $T(z,\rho)$ were summarized by \citet{GNT75}, cf. also \citet{M08}. We note two alternative forms for $T(z,\rho)$ given by \citet{GNT75}:
\bea
T(z,\rho)&=&e^{-\rho\gamma_\phi}\ln\frac{1-z}{1+z}
+\int_{-1}^1\frac{{\rm d}\beta}{\beta-z}\left(e^{-\rho\gamma}-e^{-\rho\gamma_\phi}\right),
\nn
\\
T(z,\rho)&=&-2\rho\int_0^z\frac{{\rm d} x}{[(1-x^2)(1-z^2)]^{1/2}}
K_1\left[\left(\frac{1-x^2}{1-z^2}\right)^{1/2}\rho\right]
+i\pi e^{-\rho\gamma}.
\label{GNTa}
\eea
In our detailed calculations we compared all three forms, and confirmed their equivalence.

\begin{figure}
	\begin{minipage}[ht]{1.0\linewidth}\centering
		\psfragfig[width=1.0\textwidth]{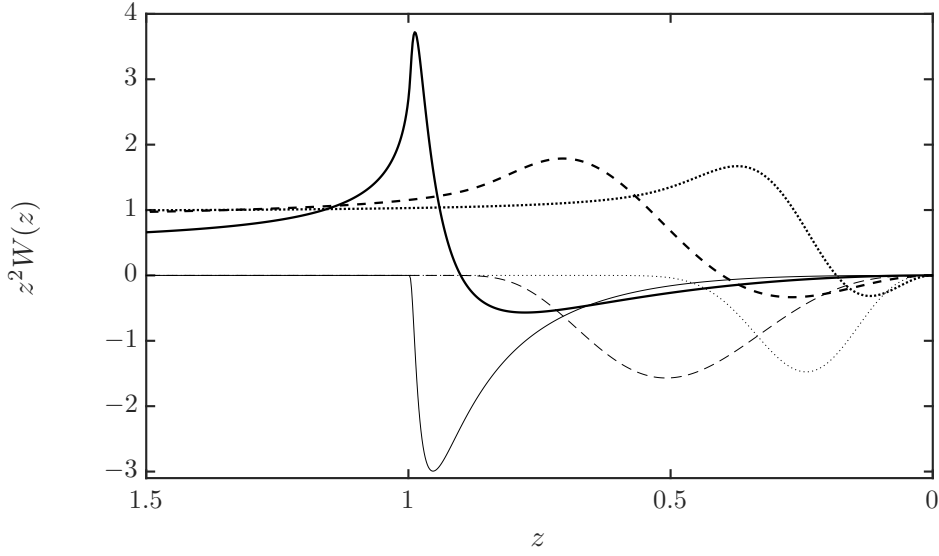}
	\end{minipage}
	\caption{The RPDF $z^2W(z)$ is plotted as a function of $z$ for 1D J\"uttner distributions. The thick curves correspond to the real part and the thin curves to the imaginary part of $ z^2W(z) $ for $\rho=50$ (dotted), $\rho=10$ (dashed) and $\rho=1$ (solid). The imaginary parts are identically zero for $z\ge1$ and negative for $z<1$. Note that $z$ increases from right to left to facilitate comparison with dispersion curves shown below.}
\label{fig:MGKF3a}
\end{figure}

\begin{figure}
\begin{minipage}[ht]{1.0\linewidth}\centering
    \psfragfig[width=0.49\columnwidth]{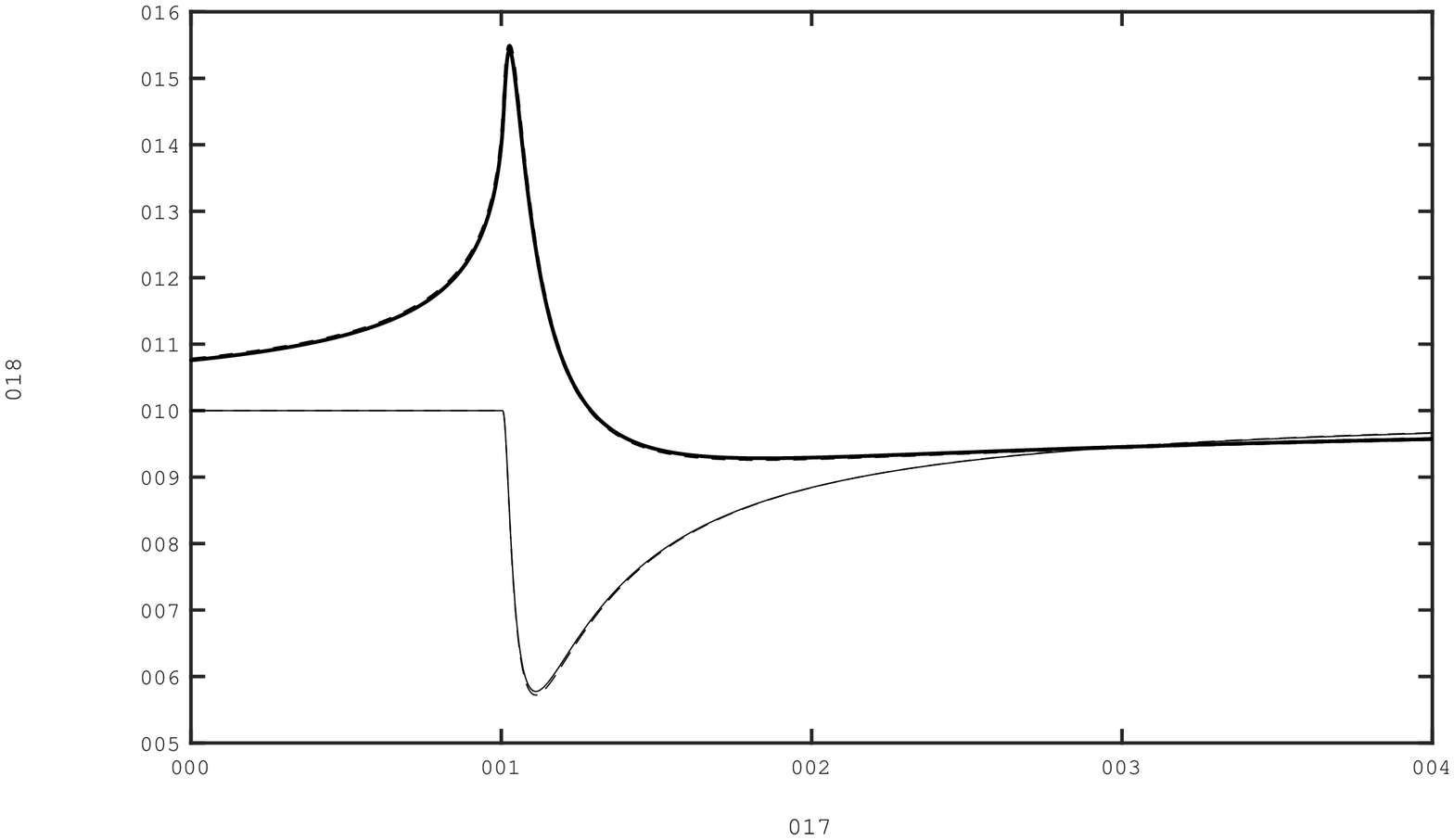}
    \psfragfig[width=0.49\columnwidth]{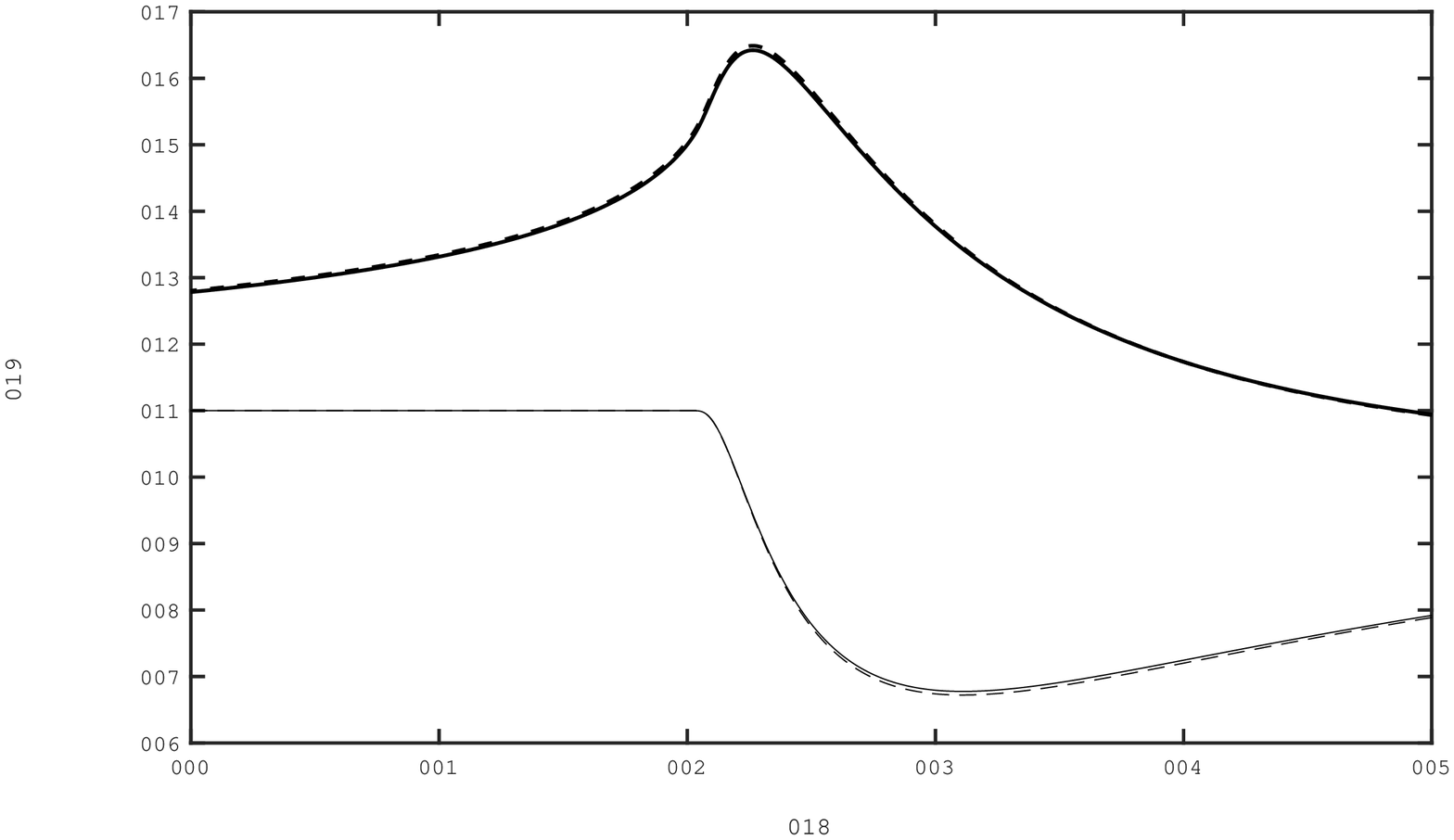}
\end{minipage}
\caption{As for Figure~\ref{fig:MGKF3a} but with $ \rho z^2 W(z) $ plotted against $ (1-z)/\rho^2 $ for $\rho=0.1$ (dashed) and  $\rho=0.01$ (solid) around $z=1$ on a fine scale (left) and on a very fine scale (right). With these scalings of the vertical and horizontal axes, the plots for $ \rho = 0.1 $ and $ \rho = 0.01 $ are nearly indistinguishable.}
\label{fig:RPDF1}  
\end{figure}

Examples of $z^2W(z)$ for the distribution (\ref{Juttner}) are shown in Figure~\ref{fig:MGKF3a} for three temperatures, ranging from a nonrelativistic value, $\rho=50\gg1$ (dotted), to a value, $\rho=1$ (solid), where relativistic effects are significant. A similar plot was presented by \citet{MG99}, and we make two notable changes in Figure~\ref{fig:MGKF3a}; we include the imaginary parts, shown by the thin curves, and we plot the curves such that the superluminal regime, $z>1$, is to the left of $z=1$ and the subluminal regime is to the right of $z=1$. The peak in the RPDF evident in Figure~\ref{fig:MGKF3a} becomes higher, narrower and closer to $z=1$ with decreasing $\rho$. We show this peak on a fine scale and on a very fine scale in Figure~\ref{fig:RPDF1} for $ \rho = 0.1 $ (dashed) and $ \rho = 0.01 $ (solid). Note that the shape of the peak near $z=1$ scales in a characteristic way with $\rho$. Numerical estimates based on the scaling apparent in these figures are given below.

For numerical results and plots we use, unless stated otherwise, pulsar period $ P = 1~{\rm s} $, period derivative $ \dot{P} = 10^{-15} $, emission height at radius $ r/r_{\rm L} = 0.1 $ where $ r_{\rm L} = Pc/2\pi $ is the light cylinder radius, and multiplicity $\kappa = 10^5 $. For these parameter values $ \beta_{\rm A} \approx 5.0\times10^3 $ \& $ \gamma_{\rm A} \approx 3.6\times10^3 $ for $ \rho = 0.1$, and $ \beta_{\rm A} \approx 1.6\times10^3 $ \& $ \gamma_{\rm A} \approx 1.1\times10^3 $ for $ \rho = 0.01 $, where $ \gamma_{\rm A} $ is given by~\eqref{gammaA}. In some calculations we vary $ \beta_{\rm A} $ which can be achieved by varying a number of pulsar parameters as
\begin{equation}
    \beta_A^2 
        \approx 2.6\times10^7\left(\frac{10}{\langle\gamma\rangle}\right)\left(\frac{10^5}{\kappa}\right)\left(\frac{\gamma_s}{10^3}\right)\left(\frac{{\dot P}/P^3}{10^{-15}}\right)^{1/2}\left(\frac{r/r_L}{0.1}\right)^{-3}.
\end{equation}

\subsection{Properties of the RPDF}
\label{sec:RPDF_properties}

\begin{figure}
	\begin{minipage}[ht]{1.0\linewidth}\centering
		\psfragfig[width=0.9\textwidth]{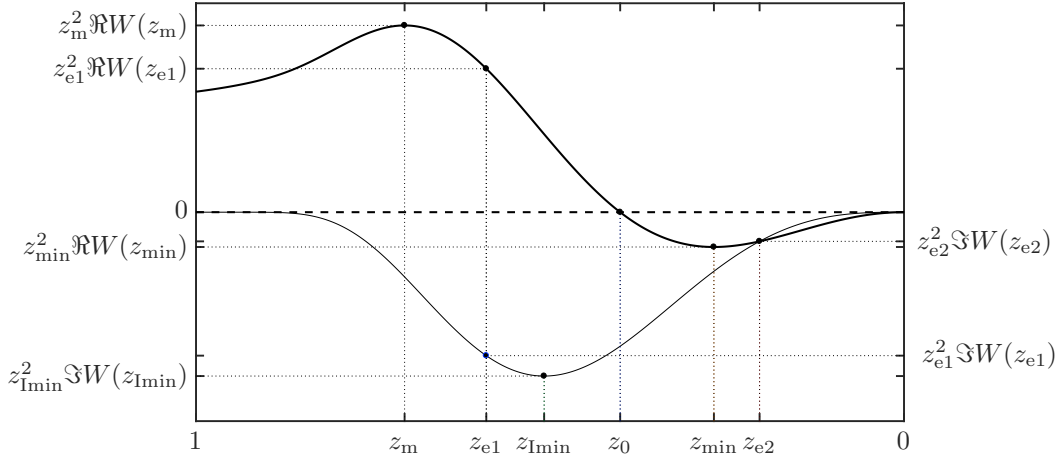}
	\end{minipage}
	\caption{A plot of $ z^2 \Re W(z) $ (thick solid) and $ z^2 \Im W(z) $ (thin solid) over $ 1 > z > 0 $ illustrating specific values of $ z $: $ z_{\rm m} $ where $ z^2\Re W(z) $ is a maximum, $ z_0 $ where $ z^2 \Re W(z) $ passes through zero, $ z_{\rm min} $ where $ z^2 \Re W(z) $ is a minimum, $ z_{\rm Imin} $ where $ z^2 \Im W(z) $ is a minimum, and $ z_{\rm e1,2} $ where $ \left|z^2\Re W(z)\right| = \left|z^2\Im W(z)\right| $.}
\label{fig:schematic}
\end{figure}

\begin{figure}
\begin{minipage}[ht]{1.0\linewidth}\centering
    \psfragfig[width=1.0\columnwidth]{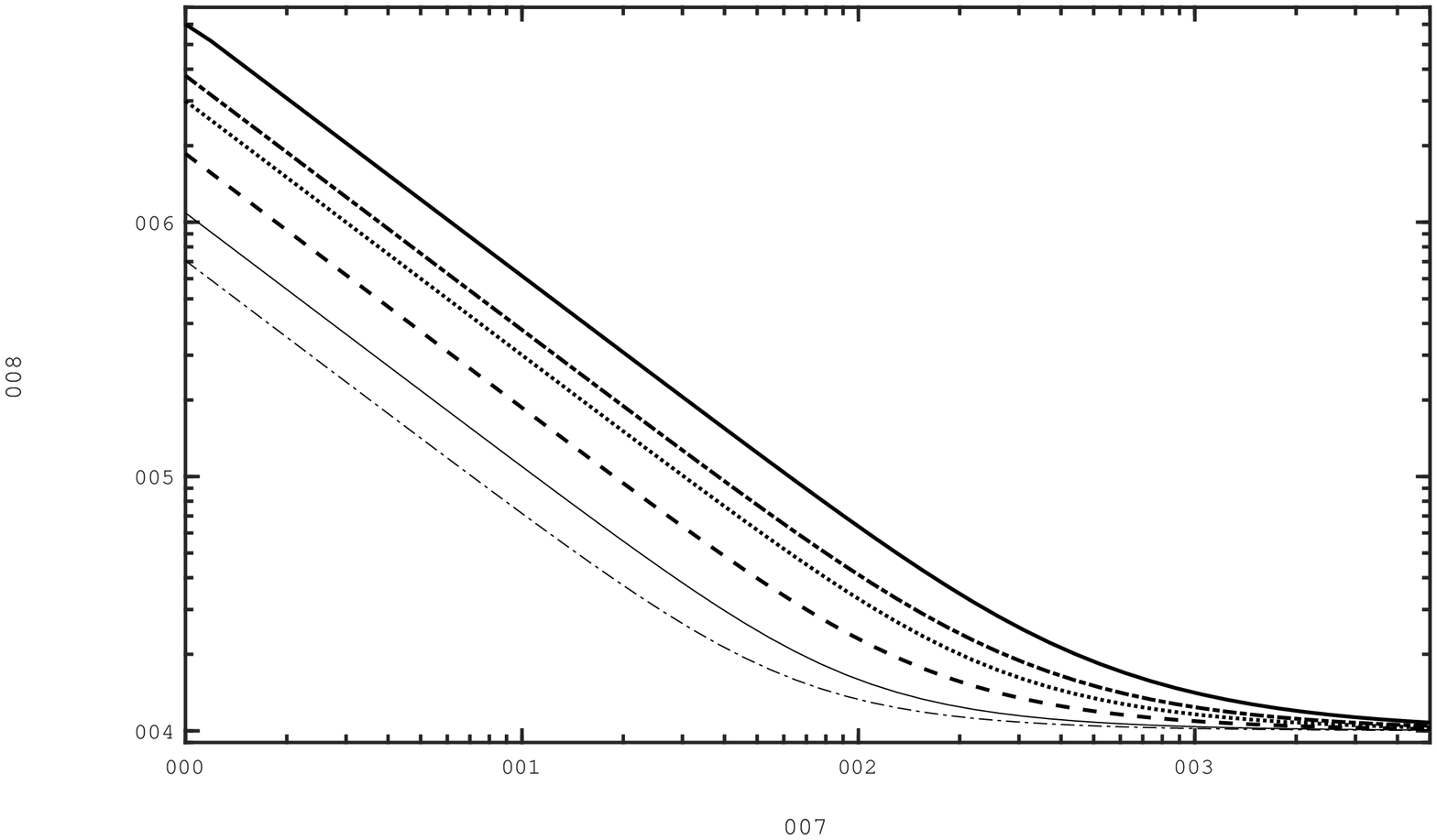}\\\vspace{3mm}
    \psfragfig[width=1.0\columnwidth]{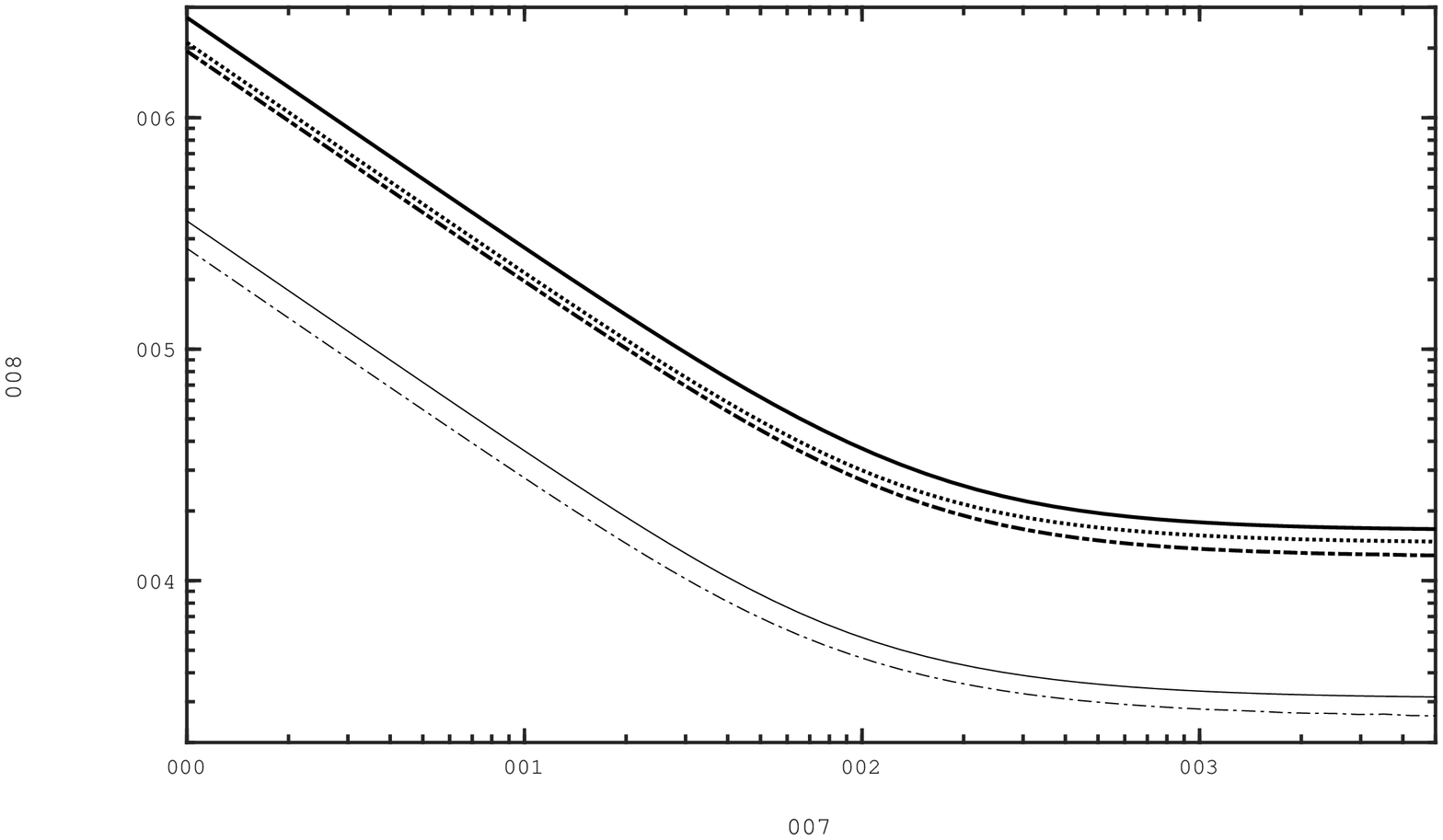}
\end{minipage}
    \caption{TOP: plots of $ \gamma_\phi $ against $ \rho $ with $ \gamma_\phi $ evaluated at $ z = z_{\rm m} $ (thick solid), at $ z = z_0 $ (dashed), at $ z = z_{\rm min} $ (thin solid), at $ z = z_{\rm Imin} $ (dotted), at $ z = z_{\rm e1} $ (thick dash-dotted) and at $ z = z_{\rm e2} $ (thin dash-dotted). BOTTOM: magnitude of $ \Re z^2W(z) $ at $ z = z_{\rm m} $ (thick solid), at $ z = z_{\rm min} $ (thin solid), at $ z_{\rm e1} $ (thick dash-dotted) and $ z_{\rm e2} $ (thin dash-dotted) where $ \left|\Im z^2W(z)\right| = \left|\Re z^2W(z)\right| $, and at $ z = z_{\rm Imin} $ where $ \left|\Im z^2W(z)\right| $ (dotted).}
\label{fig:z2W_info}  
\end{figure}

In Figure~\ref{fig:schematic} we present a typical plot of the real (thick solid) and imaginary (thin solid) part of the RPDF $ z^2 W(z) $ over $ 1 > z > 0 $. We denote the critical $ z $ values and the corresponding value of the real and imaginary parts of $ z^2W(z) $ at these $ z $ values. These critical values of $ z $ are $ z_{\rm m} $ where $ z^2\Re W(z) $ is a maximum, $ z_0 $ where $ z^2 \Re W(z) $ passes through zero, $ z_{\rm min} $ where $ z^2 \Re W(z) $ is a minimum, $ z_{\rm Imin} $ where $ z^2 \Im W(z) $ is a minimum, and $ z_{\rm e1,2} $ where $ \left|z^2\Re W(z)\right| = \left|z^2\Im W(z)\right| $. We note that
\begin{equation}\label{eq:critical_z_order}
    0 < z_{\rm e2} < z_{\rm min} < z_0 < z_{\rm Imin} < z_{\rm e1} < z_{\rm m} < 1.
\end{equation}
In the rest frame of the plasma these critical points only depend on the value of $ \rho $. For $ \rho \ll 1 $, the critical $ z $ values vary as $ z \approx 1 - \alpha_1 \rho^2 $ and the corresponding Lorentz factors as $ \gamma_\phi \approx \alpha_2/\rho $, and the real and imaginary components of $ z^2W(z) $ vary as $ \approx \alpha_3/\rho$. We give approximate values of $ \alpha_i $ in Table~\ref{tab:critical_values} for both $ \rho \ll 1 $ and $ \rho = 1 $. We plot these critical points in Figure~\ref{fig:z2W_info} where $ \rho $ varies between $ 10^{-2} $ and 50.

\begin{table}
\centering
\caption{Empirical values for parameters $ \alpha_i $ for $ i = 1, 2, 3 $.}

\setlength\tabcolsep{3mm}
\bgroup
\def\arraystretch{1.5}%
\begin{tabular}{l|l|llllll} 
\multicolumn{2}{r}{Value of $ \alpha_i $ at:}       & $ z_{\rm m} $     & $ z_{\rm e1} $    & $ z_{\rm Imin} $  & $ z_{0} $ & $ z_{\rm min} $   & $ z_{\rm e2} $  \\ 
\cline{1-8}
\multirow{3}{*}{$ \rho \ll 1 $}     & $ \alpha_1 $  & 0.0132            & 0.0351            & 0.0555            &  0.144    & 0.421             & 0.995\\
                                    & $ \alpha_2 $  & 6.15              & 3.77              & 3.00              & 1.86      &  1.09             & 0.710  \\ 
                                    & $ \alpha_3 $  & 2.73              & 1.95              & -2.12             & 0         & -0.361            & -0.275  \\
\cline{1-8}
\multirow{3}{*}{$ \rho = 1 $}       & $ \alpha_1 $  & 0.0124            & 0.0301            & 0.0469            & 0.0997    & 0.223             & 0.342  \\
                                    & $ \alpha_2 $  & 6.36              & 4.11              & 3.30              & 2.30      & 1.59              & 1.33 \\
                                    & $ \alpha_3 $  & 3.72              & 2.71              & -2.99             & 0         & -0.568            & -0.463  \\
\cline{2-8}
\end{tabular}
\egroup
\vspace{3mm}
\label{tab:critical_values}
\end{table}

The average value of powers of $ \gamma $ over a J\"uttner distribution are related to $ \rho $ as shown in Table~\ref{tab:averages} \citep{MG99}. The approximations given for $ \rho \ll 1 $ are particularly important and simple.

\begin{table}
\centering
\caption{Averages over a J\"uttner distribution as given by~\cite{MG99}. The function $ K_i(\rho) $ are modified bessel functions of second type, and $ Ki_n(\rho) $ are Bickley functions defined as the $n^{\rm th} $ integral of $ K_0(\rho) $.}

\setlength\tabcolsep{3mm}
\bgroup
\def\arraystretch{1.5}%
\begin{tabular}{lcc}
    Average                     & Exact Value & Approximation for $ \rho \ll 1 $  \\ 
    \hline
    $ \av{\gamma} $             & $ \frac{K_2(\rho) + K_0(\rho)}{2K_1(\rho)} $ & $ \frac{1}{\rho} $\\
    $ \av{\gamma^n} $           & $ \frac{K_{n+1}(\rho)}{2^n K_1(\rho)} + \sum_{i = 1}^n\frac{(n + 2 - i)K_{n - i}(\rho)}{2^n K_1(\rho)}$ & $ \frac{n!}{\rho^n} $\\
    $ \av{1/\gamma} $           & $ \frac{K_0(\rho)}{K_1(\rho)} $ & $ \rho\left[\ln(2/\rho) - 0.577\ldots\right] $  \\
    $ \av{1/\gamma^{n + 1}} $    & $ \frac{Ki_n(\rho)}{K_1(\rho)} $ & $ \frac{\sqrt{\pi}\Gamma(n/2)\rho}{2\Gamma(n/2 + 1/2)} $  
\end{tabular}
\egroup
\label{tab:averages}
\end{table}

\section{Wave dispersion for $\rho\ge1$}
\label{sect:cold}

In this section, before considering wave dispersion in a pulsar plasma with $\rho \ll 1$, for comparison we discuss nonrelativistic counterparts: the cold-plasma limit, $\rho\to\infty$ and a nonrelativistic thermal case $\rho\gg1$. We then discuss the case $\rho=1$. The highly relativistic regime, $\rho\ll1$, is discussed in the next section.

\subsection{Cold plasma limit}

The cold plasma limit corresponds to
\begin{equation}\label{eq:cold_plasma_condition}
    \rho \to \infty, \quad
    \av{\gamma} \to 1,\quad {\rm and}\quad
    z^2W(z)\to1.
\end{equation}
The nonzero terms of $\Lambda_{ij}$ are then given by equation (\ref{Lambdaij1}) with $a=1+1/\beta_{\rm A}^2$, $b=1$ and $z^2W(z)=1$. This limit may also be treated using the cold-plasma model, with the ions replaced by positrons. In cold-plasma theory, it is conventional to solve the dispersion equation for the square of the refractive index, $N^2=k^2c^2/\omega^2=1/z^2\cos^2\theta$, as a function of $\omega$ and the angle $\theta$ of wave propagation \citep{Stix62}.

\subsubsection{The X~mode}

The dispersion equation for the X-mode in the cold plasma limit follows from~\eqref{eq:oblique_X_mode} using~\eqref{eq:cold_plasma_condition} as
\be \label{Xdr}
    z^2 = \frac{z_{\rm A}^2}{\cos^2\theta},
    \quad\text{giving}\quad
    N^2 = \frac{1}{z_{\rm A}^2}=1 + \frac{1}{\beta_{\rm A}^2}.
\ee
The X~mode may be interpreted as a magnetoacoustic wave with $z_{\rm A}c$ the MHD speed when the displacement current is included. The polarization vector for the X~mode is along the $2$-axis, that is, along the direction ${\bi k}\times{\bi B}$.

\subsubsection{Parallel A~and L~modes}

With the cold plasma assumption~\eqref{eq:cold_plasma_condition} one obtains from~\eqref{eq:parallel_modes} the expression for A~mode in a cold plasma as 
\be \label{Apar}
    z^2 = z_{\rm A}^2,
    \quad\text{implying}\quad
    N^2 = \frac{1}{z_{\rm A}^2}.
\ee
From~\eqref{Xdr} and~\eqref{Apar} it is evident that the dispersion relations for the X~and A~modes are the same for parallel propagation. The polarization of the A~mode is along the $1$-axis. 

The expression for the L~mode in a cold plasma is obtained from~\eqref{eq:parallel_modes} using~\eqref{eq:cold_plasma_condition} as
\be \label{Lpar}
    \omega^2 = \omega_{\rm p}^2.
\ee
The polarization is longitudinal, which is along ${\bi B}$ for parallel propagation.

\begin{figure}
	\begin{minipage}[ht]{1.0\linewidth}\centering
		\psfragfig[width=1.0\textwidth]{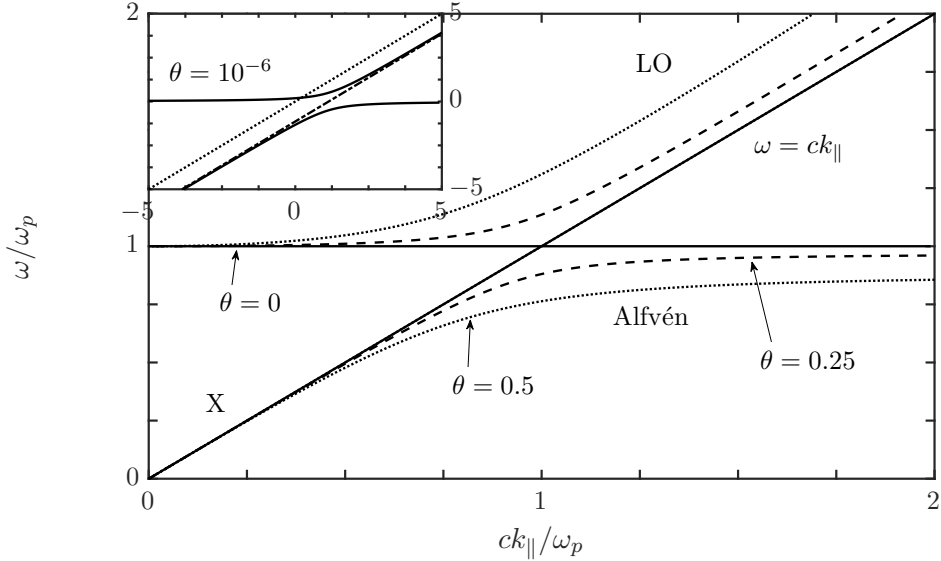}
	\end{minipage}
	\caption{Wave properties in the cold limit. The axes of the inset are $ (\omega/\omega_{\rm p} - 1)\times 10^{6} $ and $ (ck_\parallel/\omega_{\rm p} - 1)\times10 ^{6} $. The dotted line in the inset is the light line, $ z = 1 $, and the dashed line is the X mode.}
\label{fig:RPDF}
\end{figure}

\subsubsection{Alfv\'en and O~modes}

The Alfv\'en and O~modes~\eqref{eq:Alfven_O_modes} using the cold plasma limit~\eqref{eq:cold_plasma_condition} reduces to
\begin{equation}\label{eq:Alfven_O_modes_cold}
    \omega^2 
        = \frac{(z^2 - z_{\rm A}^2)\,\omega_{\rm p}^2}{z^2 - z_{\rm A}^2 - \tan^2\theta}
    \quad{\rm or}\quad
    z^2 
        = z_{\rm A}^2 + \frac{\omega^2\,\tan^2\theta}{\omega^2 - \omega_{\rm p}^2}.
\end{equation}
For $\beta_{\rm A}^2 \gg 1 $ we have $ z_{\rm A}^2 \approx 1 $ which allows us to write~\eqref{eq:Alfven_O_modes_cold} as
\be \label{colddrLO}
    N^2 
        \approx \frac{\omega^2-\omega_{\rm p}^2}{\omega^2-\omega_{\rm p}^2\cos^2\theta}
        \approx 1-\frac{\omega_{\rm p}^2}{\omega^2}\sin^2\theta,
\ee 
where the final approximation applies for $\omega^2\gg\omega_{\rm p}^2$. Equation (\ref{colddrLO}) implies propagating waves for $\omega^2>\omega_{\rm p}^2$ and for $\omega^2<\omega_{\rm p}^2\cos^2\theta$, with a stop band (evanescent waves) in the range $\omega_{\rm p}^2\cos^2\theta<\omega^2<\omega_{\rm p}^2$. The higher frequency branch corresponds to the O~mode  and the lower frequency branch to the Alfv\'en mode. For this cold plasma case the reconnection of the L~and A~modes to form these two oblique modes is illustrated in Figure~\ref{fig:RPDF}, which is similar to a figure presented by \citet{L99}.

In a relativistic plasma, an approximation to~\eqref{eq:Alfven_O_modes} that is similar to (\ref{colddrLO}) may be obtained by regarding $\omega_{\rm L}^2(z)$ as a constant (over a small range of $ z $ values) and assuming $b\approx1$, $z_{\rm A}^2\approx1$. The dispersion relation reduces to
\be
N^2\approx\frac{\omega^2-\omega_{\rm L}^2(z)}{\omega^2-\omega_{\rm L}^2(z)\cos^2\theta}.
\label{colddrLO2}
\ee 
 Analogous to the cold-plasma case, as $\theta$ increases the reconnected modes move apart, the O~mode to higher $\omega$ and larger $z$, and the Alfv\'en mode to lower $\omega$. 

\begin{figure}
	\begin{minipage}[ht]{1.0\linewidth}\centering
		\psfragfig[width=1.0\textwidth]{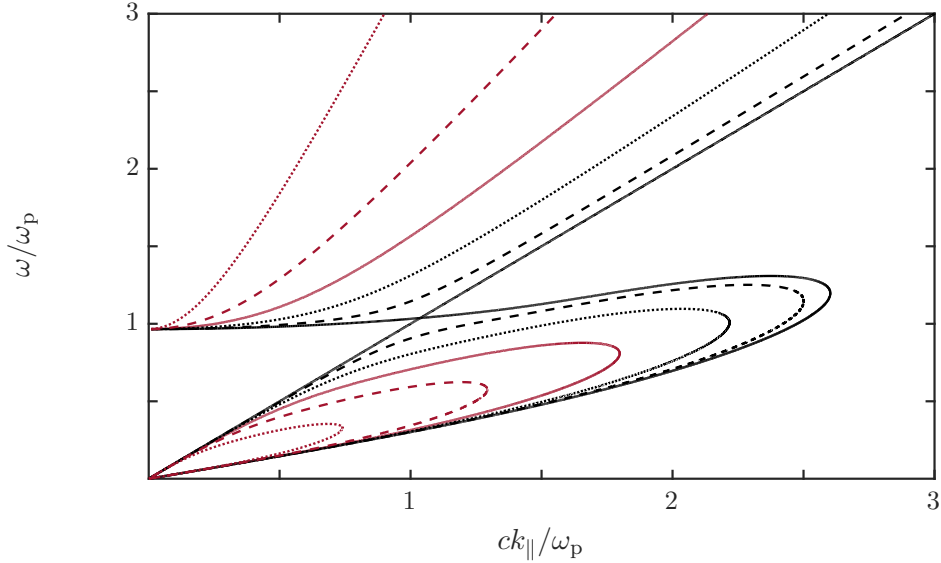}
	\end{minipage}
	\caption{Dispersion curves for $\rho=20$. The solid black curves correspond to the L~and A~modes for $\theta=0$, and the other nested curves are for the O~and Alfv\'en modes with $\theta$ increasing in steps of $0.25\,$rad. The X~mode (not shown) is degenerate with the A~mode for $\theta=0$.}
	\label{fig:dispersion_rho_20_th_0t125} 
\end{figure}

\subsubsection{Crossover frequency}

The dispersion relations for the A~and L~modes for strictly parallel propagation cross each other at the crossover frequency
\be \label{omegaco}
    \omega = \omega_{\rm L}(z_{\rm A}) \equiv \omega_{\rm co},\quad
    z = z_{\rm A}.
\ee
A crossover always occurs in the cold-plasma limit, for which the dispersion relations reduce to the horizontal line, $\omega=\omega_{\rm p}$ (L~mode), and the oblique line, $z=z_{\rm A}$ (A and X~modes) that passes through the origin in the $\omega$-$z$ plane.

\subsection{Dispersion curves for $\rho=20$} 

The transition from the cold-plasma limit, $\rho\to\infty$, to the highly relativistic limit, $\rho\ll1$, leads to a dramatic change in the wave properties. To follow how this transition occurs, it is helpful to consider two intermediate cases: one where thermal effects are significant and relativistic effects are small, and another where relativistic effects are important. We choose the intermediate cases $\rho=20$ and $\rho=1$.  

\subsubsection{L mode for $\rho=20$} 

The dispersion curves are illustrated for $\rho=20$ in Figure~\ref{fig:dispersion_rho_20_th_0t125}. The dispersion relation for the L~mode changes from the horizontal line $\omega=\omega_{\rm p}$ in a cold plasma to the solid black curve in Figure~\ref{fig:dispersion_rho_20_th_0t125}, which may be regarded as a plot of $\omega/\omega_{\rm p}$ versus $ ck_\parallel/\omega_{\rm p} = (\omega/\omega_{\rm p})/z$. The cutoff frequency, $ \omega_x = \omega_{\rm L}(\infty) = \av{1/\gamma^3}\omega_{\rm p} $, moves to just below $\omega_{\rm p}$ due to a relativistic correction $ \av{1/\gamma^3} $. The frequency increases with increasing $k_\parallel$, crosses the light line at $\omega_1 = \omega_{\rm L}(1) = (2\av{\gamma} - \av{1/\gamma})\omega_{\rm p} $, $k_\parallel c=\omega_1$ and reaches its maximum at $z=z_{\rm m}$, corresponding to $\omega/\omega_{\rm p}\approx1.5$ and $ck_\parallel/\omega_{\rm p}=2.6$. 

The L~mode is double valued, with a higher-frequency portion and a lower-frequency portion joining at what we refer to as a turnover. The turnover, as a function of $z$, occurs at $z=z_{\rm m}$, where the RPDF has its maximum. The lower-frequency portion of the dispersion curve extends to $\omega=0$, which is approached along the line $z=z_0$ (or $ \omega/\omega_{\rm p} = z_0 ck_\parallel/\omega_{\rm p} $), where the RPDF passes through zero. The higher-frequency portion may be interpreted as a counterpart of the parallel Langmuir mode in a nonrelativistic, magnetized plasma; the L~mode dispersion relation~\eqref{eq:parallel_modes} may be approximated by $\omega_{\rm L}^2(z)=\omega_{\rm p}^2+3k_\parallel^2V^2$ with $\rho=c^2/V^2\gg1$. The lower-frequency portion of the dispersion curve is in a region of strong Landau damping, and we do not discuss this branch further. 

\subsubsection{Crossover of the A and L modes} 

The solid black line in Figure~\ref{fig:dispersion_rho_20_th_0t125} is the dispersion relation $z=z_{\rm A}$ for the (parallel Alfv\'en or) A~mode, which is at an angle $\psi_{\rm A}=\arctan z_{\rm A}$ to the horizontal. For the value of $\beta_{\rm A} \approx 2.1\times10^2$ chosen in the plot $\psi_{\rm A}\approx\pi/4$ and this line is indistinguishable from the light line. For this large value of $\beta_{\rm A}$ the crossover frequency cannot be distinguished from the frequency $\omega_1 = \omega_{\rm L}(1)$ in Figure~\ref{fig:dispersion_rho_20_th_0t125}. For smaller values of $\beta_{\rm A}$ the cross-over point is on the higher-frequency portion of the curve for $z_{\rm A}>z_{\rm m}$, on the lower-frequency portion of the curve for $z_{\rm m}>z_{\rm A}>z_0$, and there is no cross-over for $z_{\rm A}<z_0$. In the following discussion we assume $z_{\rm A}>z_{\rm m}$, except where we discuss the other two cases explicitly. 

\subsubsection{Oblique O and Alfv\'en modes for $\rho=20$} 

For $\tan\theta\ne0$ the L~and A~modes reconnect to form the O~mode, which moves to the upper left as $\theta$ increases, and the Alfv\'en mode, which moves to the lower right as $\theta$ increases. The turnover, corresponding to the peak in the RPDF, is in the Alfv\'en mode for $z_{\rm A}>z_{\rm m}$. The reconnected Alfv\'en mode consists of two branches, and for sufficiently small $\theta$, these are the familiar Alfv\'en mode with $z\approx z_{\rm A}$, $\omega<\omega_{\rm co}$, and a branch that follows the L~mode for $z_{\rm A}>z\gtrsim z_{\rm m}$. We refer to the latter as the turnover branch. The wave properties remain topologically similar as $\rho$ decreases, but become increasingly distorted from the mildly relativistic case $\rho=20$ as $\rho$ decreases to $\lesssim1$.

\begin{figure}
\centering
 \psfragfig[width=1.0\columnwidth]{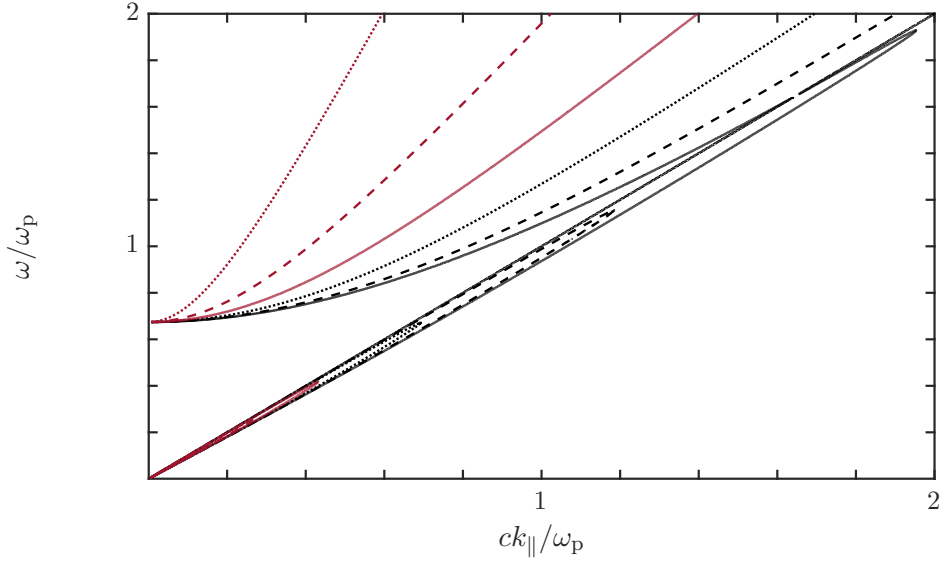}
\caption{Dispersion curves for $\rho=1$ in the same form as Figure~\ref{fig:dispersion_rho_20_th_0t125}.}
\label{fig:dispersion_rho_1_th_0t1125}  
\end{figure}

\subsection{Dispersion curves for $\rho=1$}

For the case $\rho=1$ shown in Figure~\ref{fig:dispersion_rho_1_th_0t1125}, relativistic effects are important. Comparing the dispersion curves in Figure~\ref{fig:dispersion_rho_1_th_0t1125} with those in Figure~\ref{fig:dispersion_rho_20_th_0t125}, an obvious difference is the cutoff frequency $ \omega_x = \omega_{\rm L}(\infty) $, which is only marginally below $\omega=\omega_{\rm p}$ for $\rho=20$, is significantly below $\omega=\omega_{\rm p}$ for $\rho=1$. Another notable difference is the frequency $\omega_1 = \omega_{\rm L}(1)$ at which the dispersion curve for the L~mode crosses the light line: this is marginally above $\omega=\omega_{\rm p}$ for $\rho=20$, and significantly above $\omega=\omega_{\rm p}$ for $\rho=1$. A further difference is a narrowing of the dispersion curve for the (reconnected) Alfv\'en mode, with the higher-frequency and lower-frequency portions of the dispersion curves moving closer together and closer to the light line. These intrinsically relativistic features become increasingly important as $\rho$ decreases to $\ll1$.

\section{Wave dispersion for $\rho\ll1$}
\label{sect:rhogg1}

In this section we discuss the wave properties for a pulsar plasma with $\rho\ll1$, corresponding to $\langle\gamma\rangle\approx1/\rho\gg1$.

\subsection{L~mode for $\langle\gamma\rangle\gg1$} 

Approximate forms for the dispersion relation for the L~mode, given by~\eqref{eq:parallel_modes}, for large $z$ and for $z\approx1$ were given in the early literature \citep{LM79,LMS79}. These approximations are derived by expanding $z^2W(z)$ in powers $1/z\ll1$ and in powers of $|1-z|\ll1$, respectively, and retaining only the lowest order terms:
\be
z^2W(z) \approx 
\begin{cases}
{\displaystyle
\left\langle\frac{1}{\gamma^3}\right\rangle
\left[1+\frac{3}{z^2}\left(1-\frac{\langle\gamma^{-5}\rangle}{\langle\gamma^{-3}\rangle}\right)\right]
\approx\frac{1}{\langle\gamma\rangle}
\left(1+\frac{3}{z^2}\right),
} & \text{for}\quad |z| \gg 1,\\
\ms
2\langle\gamma\rangle+(z^2-1)4\langle\gamma^3\rangle
\approx2\langle\gamma\rangle[1+12(z^2-1)\langle\gamma\rangle^2] , & \text{for}\quad |1-z^2| \ll 1,
\end{cases}
\label{Wza}
\ee
where the final forms apply for a J\"uttner distribution with $\langle\gamma\rangle\gg1$ \citep{MG99}:
\be
\langle\gamma^n\rangle=n!\langle\gamma\rangle^n,
\quad
\langle\gamma^{-1}\rangle=\langle\gamma\rangle^{-1}[\ln(2\langle\gamma\rangle)-0.577],
\quad
\langle\gamma^{-3}\rangle\approx\langle\gamma\rangle^{-1},
\quad
\langle\gamma^{-5}\rangle\approx{\textstyle\frac{2}{3}}\langle\gamma\rangle^{-1}.
\label{gammanav}
\ee
We rederive the approximations (\ref{Wza}) in Appendix~\ref{app:C} by expanding in powers of $1/z^2$ and $z-1$, respectively. Although we find that the approximation for $z^2\gg1$ is well justified, that for $z-1$ is based on an expansion that converges only for $z^2=1$. We suggest that the approximation (\ref{Wza}) for $|1-z^2|\ll1$ should not be used, at least without further justification. 

The cutoff frequency $ \omega_x $ and the frequency $ \omega_1 $ are given by
\begin{equation}
    \omega_x = \omega_L(\infty) = \av{1/\gamma^3}\omega_{\rm p},\quad
    \omega_1 = \omega_L(1) = \left(2\av{\gamma} - \av{1/\gamma}\right)\omega_{\rm p},
\end{equation}
with $ \av{1/\gamma^3} \approx 1/\av{\gamma} $ for $ \av{\gamma} \gg 1 $. For a cold plasma one obtains $ \omega_x = \omega_1 = \omega_{\rm p} $. 

The dispersion relation, $\omega=\omega_{\rm L}(z)$, just above the cutoff frequency $ \omega_x $ is then given by
\be
\omega_{\rm L}^2\approx\omega_x^2+k_\parallel^2c^2,
\label{deL1}
\ee 
which reproduces a known result \citep{LM79,LMS79,MG99}. However, analogous approximate dispersion relations for $z^2\approx1$ are questionable for the reason discussed above. Specifically, although one can evaluate $z^2W(z)$ and all its derivatives at $z^2=1$, the Taylor series in $(z^2-1)$ involving these derivatives appears to have zero radius of convergence (Appendix~\ref{app:C}).

\subsection{O~mode for $\langle\gamma\rangle\gg1$}

The O~mode has the same cutoff frequency, $\omega_x$, as the L~mode independent of $\theta$. As $\theta$ increases the dispersion curve for the O~mode deviates increasingly from that of the L~mode frequency. One may rewrite equation (\ref{eq:Alfven_O_modes}) in the form
\be
\omega^2(z,\theta)=\frac{\omega_{\rm L}^2(z)}{1+a(z)\tan^2\theta},
\qquad
a(z)=\frac{b}{z_{\rm A}^2-z^2}.
\label{az}
\ee
One has $z^2 > z_{\rm A}^2 + b\tan^2\theta$ for the O~mode, and hence $a(z)<0$, which implies that the frequency is an increasing function of $\theta$. Except for the small range $\theta^2\lesssim2/\beta_{\rm A}^2$ the dispersion curve for the O~mode is entirely in the superluminal range. The resonance condition $\beta_{\rm s}=z$, where $ \beta_{\rm s}c $ is the streaming speed, cannot be satisfied for the O~mode except for this tiny range of angles. An approximate dispersion relation for the (superluminal) O~mode follows by setting $z_{\rm A}^2\to1$, $b\to1$ in equation (\ref{az}):
\be
\omega^2_{\rm O}(z,\theta)
\approx\omega_{\rm L}^2(z)\frac{(z^2-1)\cos^2\theta}{z^2\cos^2\theta-1},
\label{omegaLO}
\ee
which applies for $z^2>1$ when $ \beta_{\rm A}^2 \gg 1 $. The frequency, $\omega_{\rm O}(1,\theta)$, at which the dispersion curve crosses the light line increases with increasing $\theta$:
\be
\omega^2_{\rm O}(1,\theta)=\omega_1^2\frac{1-z_{\rm A}^2}{1-z_{\rm A}^2-b\tan^2\theta}
\approx\frac{\omega_1^2}{1-\beta_{\rm A}^2\theta^2/2},
\label{omegaLO1}
\ee
where the approximation applies for $\beta_{\rm A}^2\gg1$, $ \av{\gamma} \gg 1 $ and $\theta^2\ll1$.

An approximate dispersion relation for the O~mode for $ z = 1/N\cos\theta \gg 1 $ and $\omega^2\gg\omega_{\rm L}^2(z = 1/N\cos\theta) $ that follows from (\ref{colddrLO2}) using~\eqref{Wza} is
 \be
    N^2_{\rm O}
        \approx \frac{1 - \omega_{\rm p}^2\sin^2\theta/\langle\gamma\rangle\omega^2}{1 + 3\omega_{\rm p}^2\sin^2\theta\cos^2\theta/\langle\gamma\rangle\omega^2}
        \approx 1 - \frac{\omega_{\rm p}^2}{\langle\gamma\rangle\omega^2}[\sin^2\theta(1 + 3\cos^2\theta)],
\label{drOmode}
\ee 
where the final expression applies for $ \omega^2 \gg (1+3\cos^2\theta) \omega_{\rm p}^2\sin^2\theta/\av{\gamma} $.

\begin{figure}
\centering
  \psfragfig[width=1.0\columnwidth]{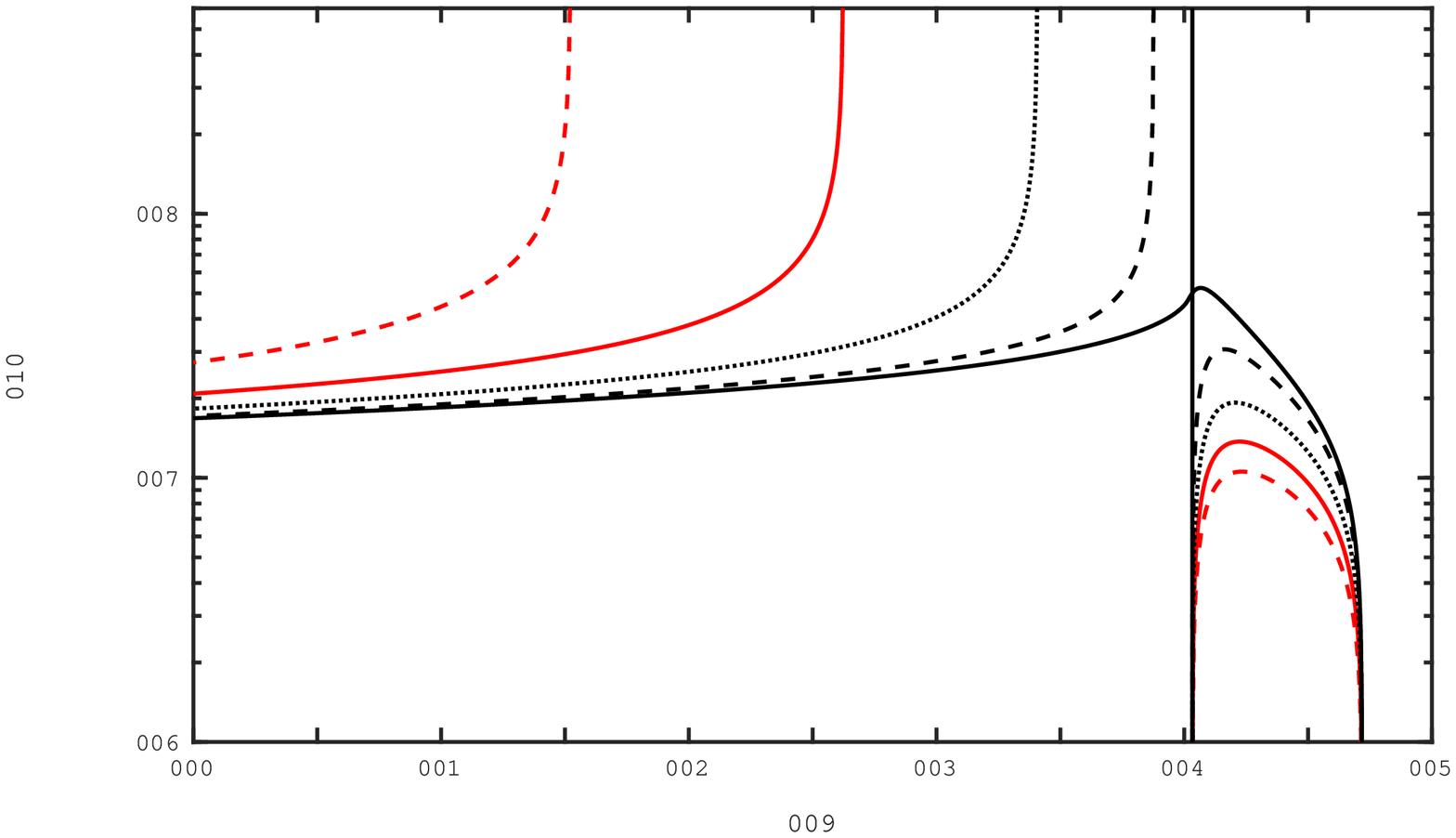}
  \vspace{0.5cm}
 \psfragfig[width=1.0\columnwidth]{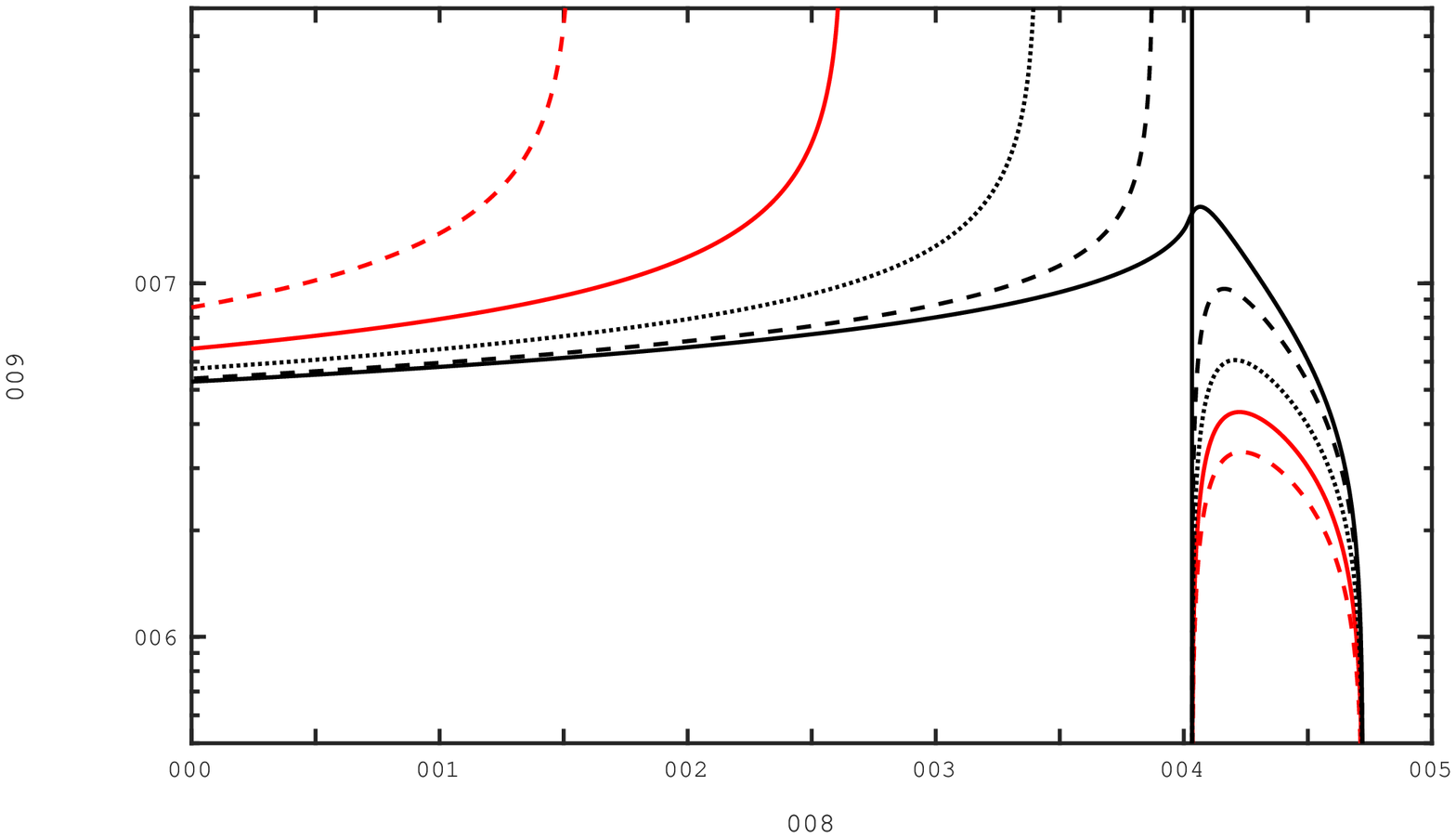}
\caption{Dispersion curves for $ \theta = 0 $ (black solid), $0.25\rho\,$rad (black dashed), $0.5\rho\,$rad (black dotted), $0.75\rho\,$rad (red solid), and $0.1\rho\,$rad (red dashed): top $\rho=0.1$ with $ \beta_{\rm A} \approx1.2\times10^2$, bottom, $\rho=0.01$ with $\beta_{\rm A}\approx1.2\times10^3$. The black solid curve corresponds to the L~mode, the solid vertical line at $ z = z_{\rm A} $ corresponds to the A~mode with the O~mode to its upper left and the Alfv\'en mode to its lower right. The Alfv\'en mode exists between $z = z_{\rm A}$, which is very close to zero in the figure with $ \gamma_{\rm A} \approx 87 $ for $\rho = 0.1$ and $ \gamma_{\rm A} = 8.7\times10^2$ for $\rho=0.01$, and  $z=z_0$. The maximum in the dispersion curve occurs near $z=z_{\rm m}$. }
\label{fig:log1-z}  
\end{figure}

\subsection{Alfv\'en mode for $\langle\gamma\rangle\gg1$}

The turnover branch of the Alfv\'en mode for $z_{\rm A}>z_{\rm m}$ and $\rho\ll1$ is dominated by the peak in $z^2W(z)$. For the case $\rho=1$ shown in Figure~\ref{fig:dispersion_rho_1_th_0t1125}, the Alfv\'en mode consists of a thin loop, with the higher-frequency and lower-frequency portions joining at a turnover that corresponds to the peak at $z=z_{\rm m}$ in the RPDF. The maximum frequency of the Alfv\'en mode decreases with increasing $\theta$ approximately along the line $z=z_{\rm m}$. For $\rho\approx1/\langle\gamma\rangle\ll1$ the loop becomes narrower with decreasing $\rho$. The format adopted in Figures~\ref{fig:dispersion_rho_20_th_0t125} and~\ref{fig:dispersion_rho_1_th_0t1125} makes it increasingly difficult to illustrate the dispersive properties due to the curves being strongly concentrated near $z=1$. 

An alternative format is used in Figure~\ref{fig:log1-z} to show the dispersion curves near $z=1$: we plot $\omega/\omega_{\rm p}$ on a log scale as a function of $(1-z)/\rho^2$ for $\rho=0.1$, with $ \beta_{\rm A} \approx 1.2\times10^2 $ and $ \gamma_{\rm A} \approx 87 $, and $0.01$, with $\beta_{\rm A} \approx 1.2\times10^3$ and $ \gamma_{\rm A} \approx 8.7\times10^2 $, where $\gamma_{\rm A}$ is defined by (\ref{gammaA}). The solid curves correspond to  the L~mode for $\theta=0$ and the solid vertical line is the dispersion curve $z=z_{\rm A}$ for the A~mode. The curves to the left of $(1-z_{\rm A})/\rho^2$ correspond to the O~mode and the curves to the right of $(1-z_{\rm A})/\rho^2$ correspond to the Alfv\'en mode. Each of $1-z_{\rm A}\approx1/2\beta_{\rm A}^2$, $1-z_{\rm m}$ and $1-z_0$ is proportional to $\rho^2\approx1/\langle\gamma\rangle^2$ as discussed in Section~\ref{sec:RPDF_properties}, so that the locations of the left asymptote, the peak and the right asymptote of the Alfv\'en mode all scale with $\rho^2$ and so coincide in the two Figures.

The Alfv\'en mode (for $\theta\ne0$) exists only for $z<z_{\rm A}$ or $\gamma_\phi< \gamma_{\rm A} \approx \beta_{\rm A}$. Its dispersion relation, given by equation (\ref{az}) with $a(z)>0$, becomes
\be
\omega^2_{\rm A}(z,\theta)
=\frac{\omega_{\rm L}^2(z)}{1+a(z)\theta^2},
\qquad
a(z)\approx\frac{\beta_{\rm A}^2\gamma_\phi^2}{\beta_{\rm A}^2-\gamma_\phi^2}.
\label{omegaA}
\ee
The Alfv\'en mode is in the range where Landau damping is nonzero. Assuming that the waves are weakly damped only for $\gamma_\phi^2\gtrsim\gamma_{\rm m}^2$, the approximations $z_{\rm m}\approx1-0.013\rho^2$, given by equation Table~\ref{tab:critical_values}, and $\langle\gamma\rangle\approx1/\rho$ imply that the waves are weakly damped for $\gamma_\phi\gtrsim6\,\langle\gamma\rangle$.

\subsection{Condition for reconnection of modes}

The dispersion curves for parallel propagation, $\omega=\omega_{\rm L}(z)$ and $z=z_{\rm A}$, do not necessarily cross in a relativistic plasma. There are three possibilities:
\begin{itemize}
\item $z_{\rm A}>z_{\rm m}$: the line $z=z_{\rm A}$ crosses the higher-frequency portion of the L~mode dispersion curve $\omega=\omega_{\rm L}(z)$ in the range $1<z<z_{\rm m}$; reconnection in this case is similar to that in the cold-plasma limit. The peak in the dispersion curve is in the (reconnected) Alfv\'en mode.
\item $z_{\rm m}>z_{\rm A}>z_0$: the line $z=z_{\rm A}$ crosses the higher-frequency portion $\omega=\omega_{\rm L}(z)$ in the range $z_{\rm m}>z_{\rm A}>z_0$.
\item $z_{\rm A}<z_0$: the line $z=z_{\rm A}$ does not cross $\omega=\omega_{\rm L}(z)$.
\end{itemize}
Simple estimates for a pulsar magnetosphere suggest very large values, $\beta_{\rm A}\gg1$, but before concentrating on this case we comment on the second and third cases.

\begin{figure}
\centering
  \psfragfig[width=1.0\columnwidth]{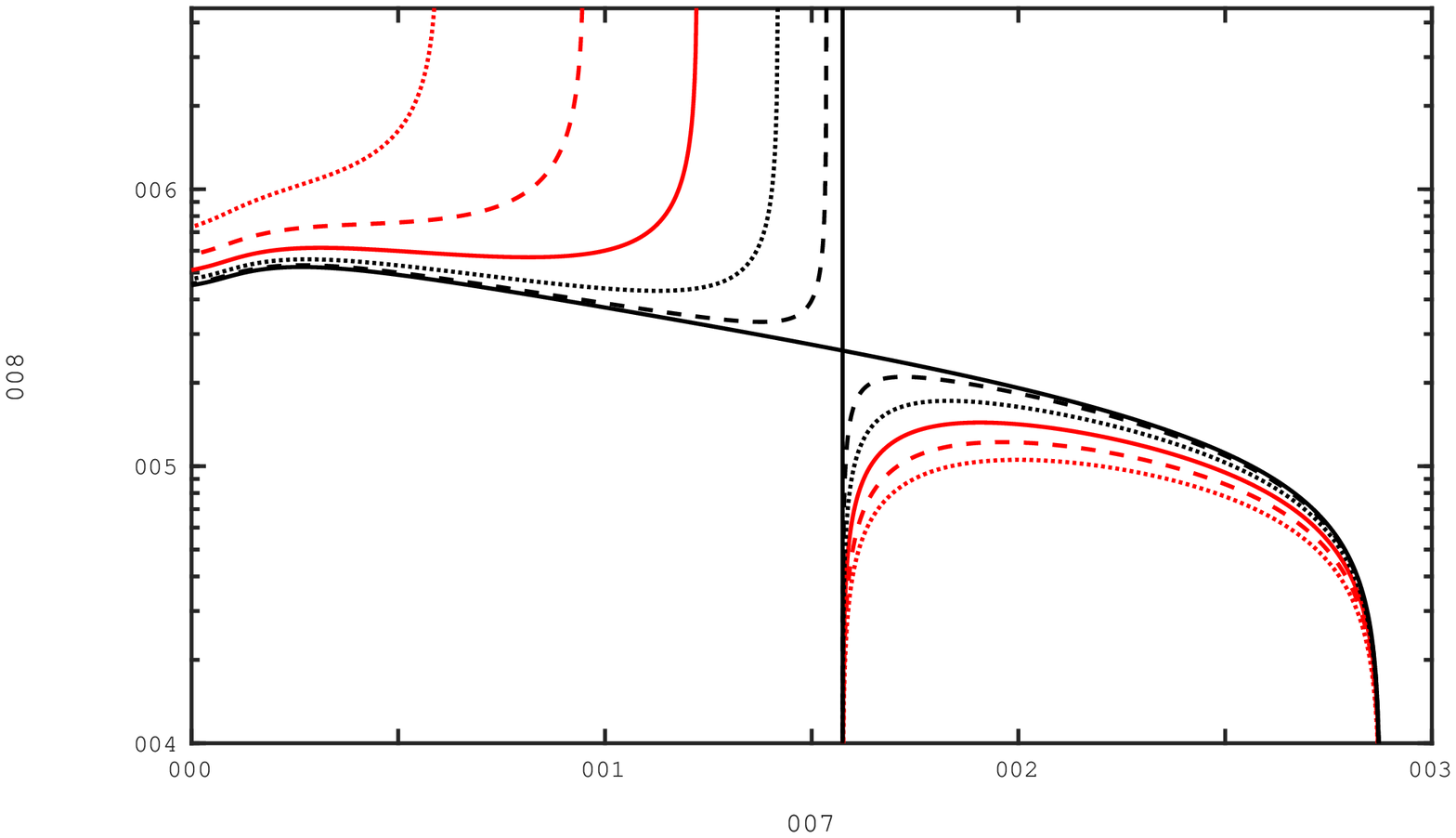}
  \vspace{0.5cm}
 \psfragfig[width=1.0\columnwidth]{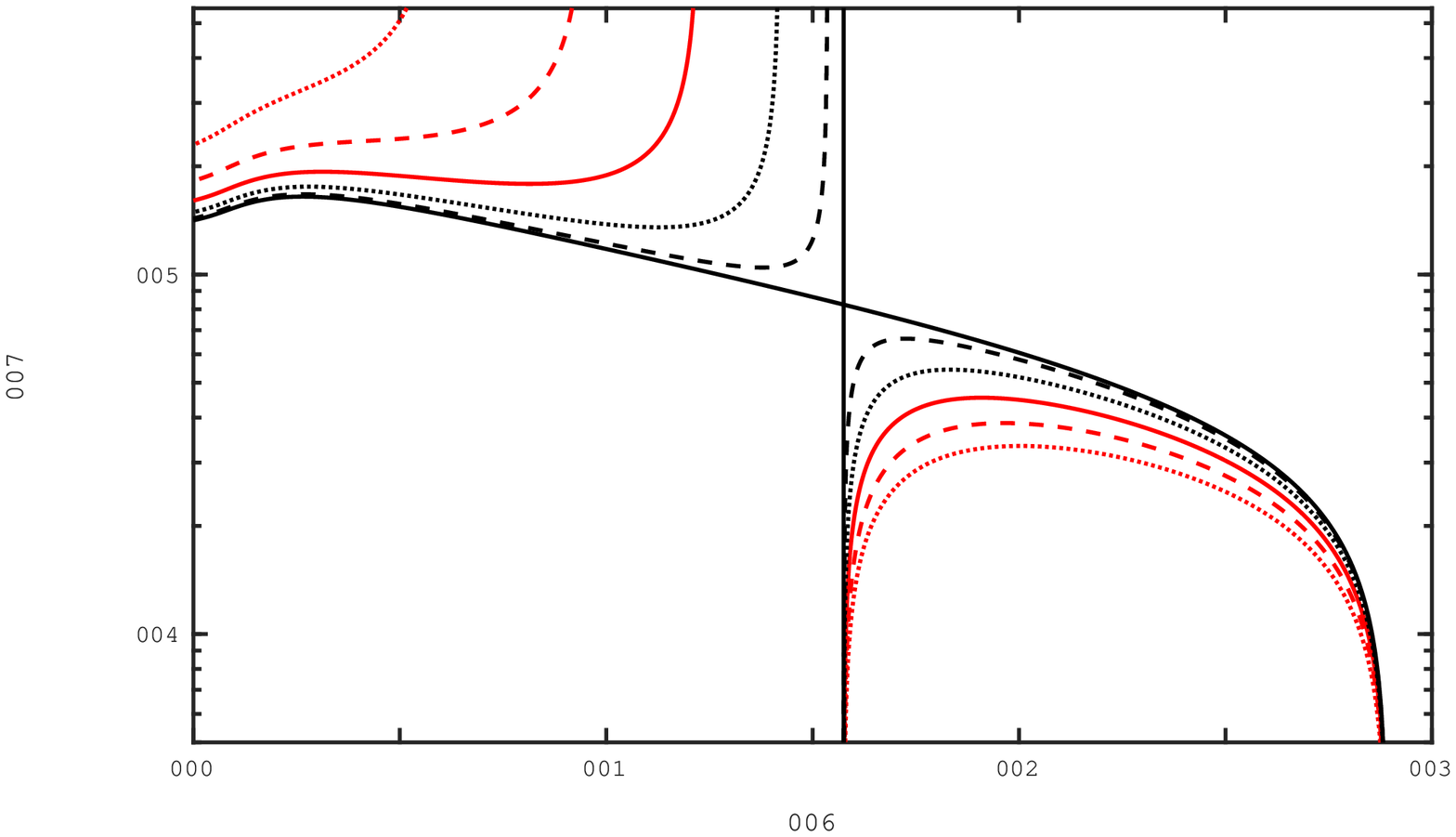}
\caption{Similar to Figure~\ref{fig:log1-z} but for $z_{\rm m}>z_{\rm A}>z_0$ and as a function of angle in steps of $0.0625\rho\,$rad: top $\rho=0.1$ with $ \beta_{\rm A} \approx 36$ and $ \gamma_{\rm A} \approx 25 $, bottom, $\rho=0.01$ with $\beta_{\rm A}\approx3.6\times10^2$ and $ \gamma_{\rm A} \approx 2.5\times10^2$.}
\label{fig:second_case}  
\end{figure}

\subsubsection{Case $z_{\rm m}>z_{\rm A}>z_0$}

Examples of the dispersion curves for $z_{\rm m}>z_{\rm A}>z_0$ are shown in Figure~\ref{fig:second_case} for $\rho=0.1$ and $\rho=0.01$. For $\theta=0$, the solid curve corresponds to the L~mode, and the solid vertical line corresponds to the A~mode at $ z = z_{\rm A} $. The maximum in the solid curve is at $z=z_{\rm m}$, to the left of $z=z_{\rm A}$. This peak leads to a maximum and a minimum in the O~mode dispersion curve for small $\theta$. The maximum and minimum become smoothed out with increasing $\theta$, and the dispersion curves are then qualitatively similar to those for $z_{\rm A}>z_{\rm m}$ when $ z_{\rm A}^2 + b\tan^2\theta > z_{\rm m}^2 $. As $\theta$ increases the Alfv\'en mode moves downward, with asymptotes at $z=z_{\rm A}$ and $z=z_0$ remaining fixed. Only the branch with $z\approx z_{\rm A}$ is Alfv\'en-like, and we refer to the branch that links the asymtotic forms as the turnover branch. Landau damping becomes important in the range $z_{\rm m}>z>z_0$, and this damping needs to be taken into account in a more detailed discussion of this case.

\begin{figure}
\centering
  \psfragfig[width=1.0\columnwidth]{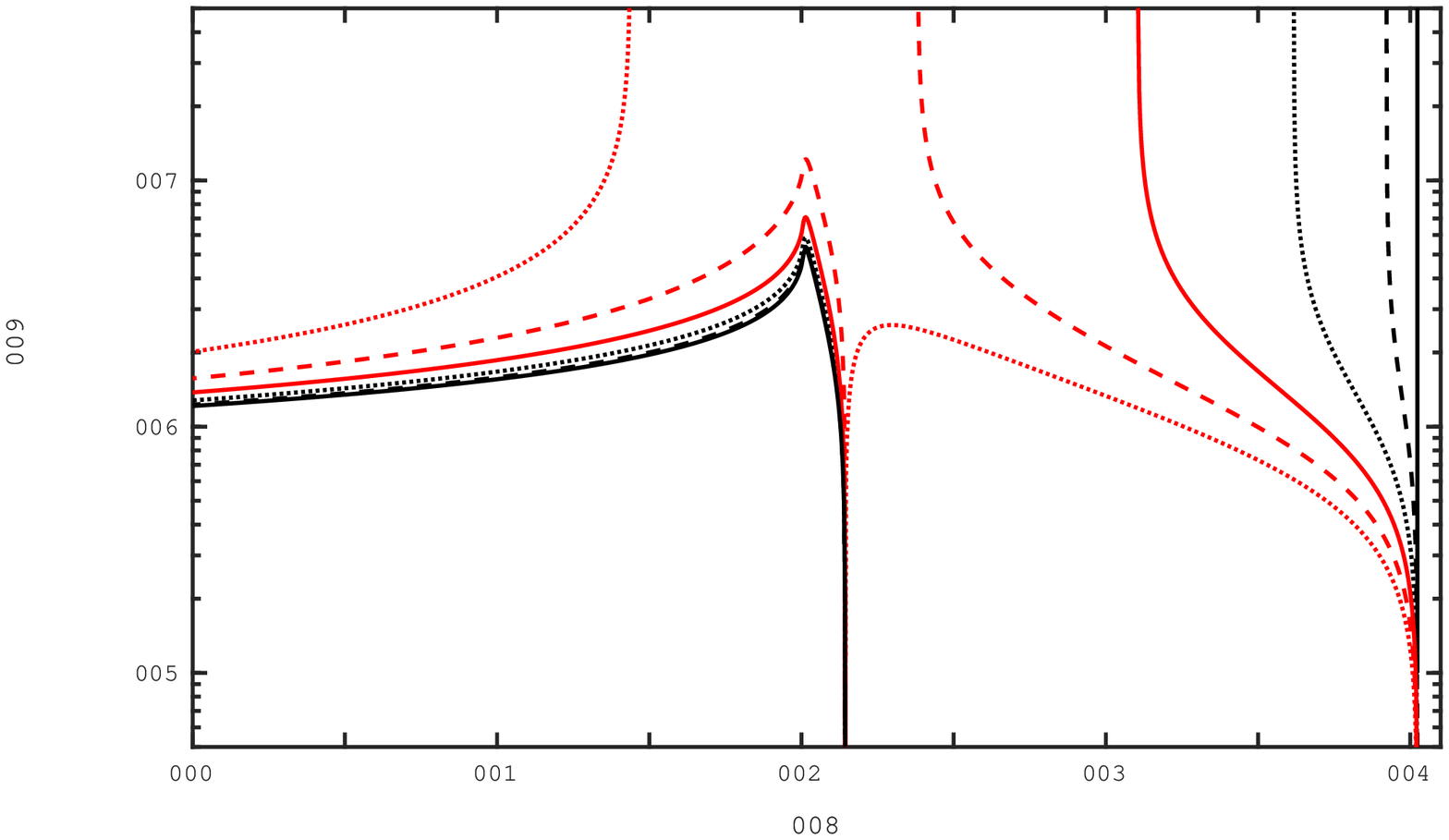}
  \vspace{0.5cm}
 \psfragfig[width=1.0\columnwidth]{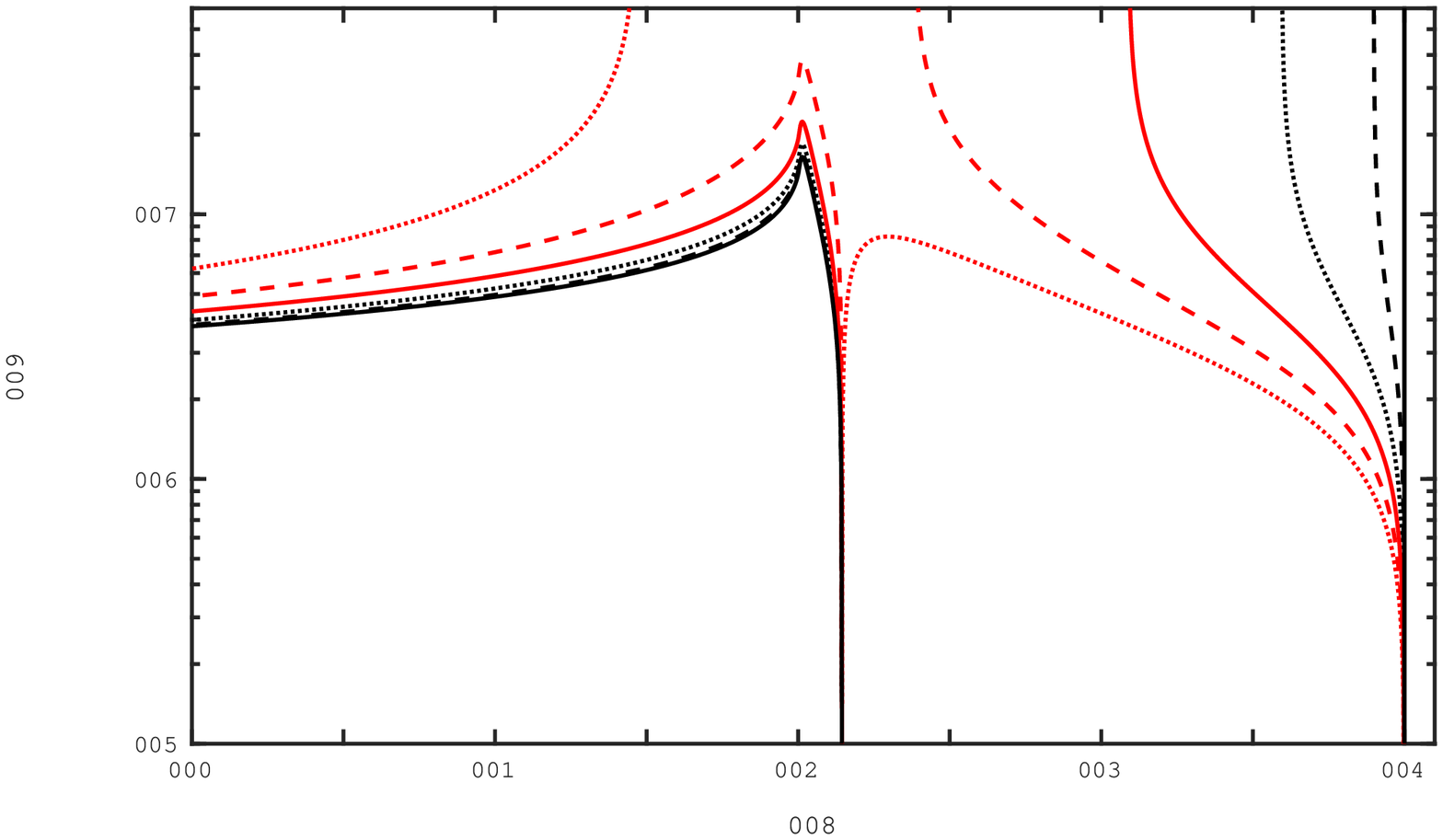}
\caption{Similar to Figure~\ref{fig:log1-z} but for $z_{\rm A}<z_0$ and as a function of angle in steps of $0.45\rho\,$rad: top $\rho=0.1$ with $ \beta_{\rm A} \approx 7.1$ and $ \gamma_{\rm A} \approx 5 $, bottom, $\rho=0.01$ with $\beta_{\rm A}\approx71$ and $ \gamma_{\rm A} \approx 50$.}
\label{fig:third_case}  
\end{figure}

\subsubsection{Case $z_{\rm A}<z_0$}

Examples of the dispersion curves for $z_{\rm A}<z_0$ are shown in Figure~\ref{fig:third_case} for $\rho=0.1$ and $\rho=0.01$. For parallel propagation the two dispersion curves do not cross. Equations (\ref{eq:Alfven_O_modes_cold}) continue to apply, with $\omega_{\rm L}^2(z)<0$. As $\theta$ increases the L~mode curve, including the peak, moves upward and to the left. The A~mode dispersion curve moves to the left. The two curves meet at $\omega=\infty$ at the angle determined by $z_0^2=z_{\rm A}^2+b\tan^2\theta$. Further increase in $\theta$ leads to the O~mode curve continuing to move upward and to the left, with the Alfv\'en mode curve moving downward, between asymptotes at $z=z_0$ and $z=z_{\rm A}$.

It is questionable whether the reconnected mode in this case should be referred to as the Alfv\'en mode. It may also be regarded as an intrinsically new mode \citep{MG99}. The argument for this is as follows. The dispersion equation for oblique propagation in the form (\ref{eq:Alfven_O_modes_cold}) gives $\omega^2$ in terms of three factors, $z^2W(z)$, $z^2-z_{\rm A}^2$ and $z^2-z_{\rm A}^2-b\tan^2\theta$. Real $\omega^2$ requires that either all three factors are positive or that one is positive and two are negative. For $z>z_0$, the O~mode corresponds to $z^2W(z)>0$ and $z^2>z_{\rm A}^2+b\tan^2\theta>z_{\rm A}^2$ and the Alfv\'en mode corresponds to $z^2W(z)>0$ and $z^2<z_{\rm A}^2<z_{\rm A}^2+b\tan^2\theta$. For $z<z_0$, with $z^2W(z)<0$ implying $\omega_{\rm L}^2(z)<0$, this additional mode exists in the range $z_0<z<z_{\rm A}$ but only for sufficiently oblique propagation, specifically, for $b\tan^2\theta>z_0^2-z_{\rm A}^2$.

\section{Properties of resonant L~mode waves}
\label{sect:Lmode}

The wave properties of the (parallel) L~mode in a pulsar plasma are of particular interest when considering beam-driven instabilities. The L~mode is the nearest counterpart to the Langmuir mode in a nonrelativistic plasma, but its properties have major differences from those of Langmuir waves. In this section we discuss some of these properties.

\begin{figure}
\begin{center}
\psfragfig[width=1.0\columnwidth]{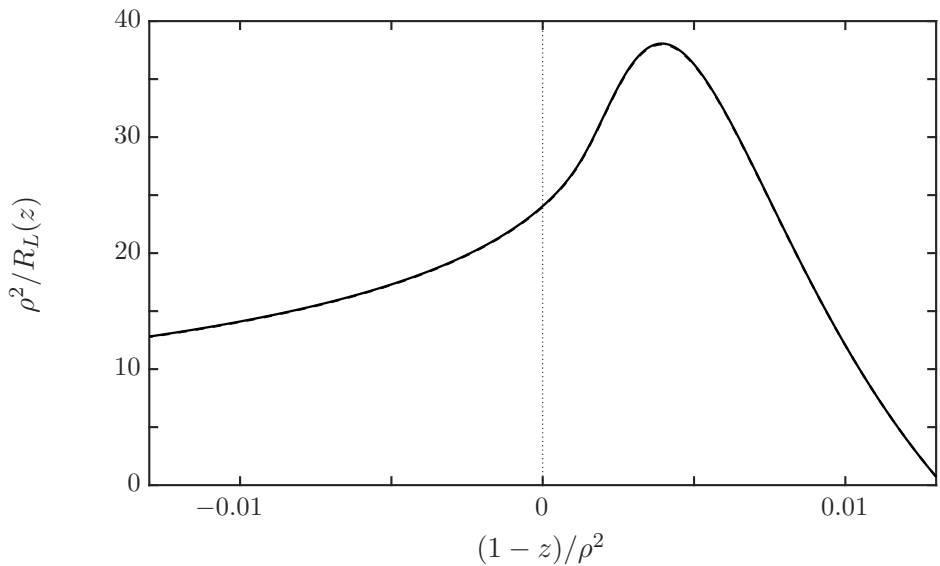}
 \caption{The function $\rho^2/R_{\rm L}(z)$ is plotted over $ (1-z)/\rho^2 $ for $ \rho = 0.1 $ (solid) and 0.01 (dashed). The two curves are indistinguishable over the scale shown. The thin dotted vertical line indicates the light line.}
 \label{fig:1/RL}  
 \end{center}
\end{figure}

\subsection{Ratio of electrical to total energy}

In general the energy in waves in a specific mode in a dispersive medium may be separated into electric, magnetic and kinetic contributions, with the kinetic contribution attributed to the perturbations in the motion of particles forced by the wave  \citep[e.g.,][Chapter 15]{1991epdm.book.....M}. The total energy, which is the sum of the three, is determined by the dispersion theory, allowing one to infer the kinetic energy contribution. The magnetic energy is zero in a longitudinal wave. The ratio of the electric to total energy, denoted as $R_{\rm L}(z) $ for the L~mode, is relevant because the rate at which work is done by any source term involves only the electric energy, but the energy that appears in waves is the total energy. 

The ratio $R_{\rm L}(z)$ is evaluated in terms of $W(z)$ in Appendix~\ref{sect:app}. Using the form (\ref{Wz}) for the RPDF, this gives
\be
R_{\rm L}(z)=-\frac{W(z)}{zdW(z)/dz}=\frac{1}{2z^2}
\left\langle\frac{z^2+\beta^2}{\gamma^3(z^2-\beta^2)^2}\right\rangle
\left\langle\frac{z^2+3\beta^2}{\gamma^3(z^2-\beta^2)^3}\right\rangle^{-1},
\label{R_L1}
\ee
which is strictly valid only for superluminal waves,  $z\ge1$. For $z\gg1$ and $z\to1$ equation (\ref{R_L1}) gives
\be
R_{\rm L}(z)\approx\frac{1}{2} - \frac{3z^2+1}{2z^2(z^2 + 3)},
\qquad
R_{\rm L}(1)\approx\frac{\langle\gamma\rangle}{4\langle\gamma^3\rangle}
\approx\frac{1}{24\langle\gamma\rangle^2},
\label{R_L2}
\ee
respectively, where the final approximation applies for a J\"uttner distribution, $\langle\gamma^n\rangle=n!\langle\gamma\rangle^n$ for $ \langle\gamma\rangle \gg 1 $. One finds that $R_{\rm L}(z)$ decreases with decreasing $z$, from $1/2$ in the limit $z\to\infty$, to $1/24\langle\gamma\rangle^2$ at $z=1$. For $z$ in the range $1\gtrsim z\ge z_{\rm m}\approx1-0.13\rho^2$ it is convenient to plot the inverse, $1/R_{\rm L}(z)$, rather than $R_{\rm L}(z)$ itself, as illustrated in Figure~\ref{fig:1/RL}. The form of $1/R_{\rm L}(z)\propto1/\rho^2\approx\langle\gamma\rangle^2$ scales with $(1-z)/\rho^2$, increasing with decreasing $z$ through its value $\approx24\langle\gamma\rangle^2$ at $z=1$ to a maximum $\approx38\langle\gamma\rangle^2$ at $z\approx1-0.004\rho^2$, and then decreasing with decreasing $z$, passing through zero at $z=z_{\rm m}$, and becoming negative in the region $z<z_{\rm m}$; as already remarked, we do not consider solutions in the region $z<z_{\rm m}$ where Landau damping is strong.

\subsection{Group speed}

The velocity of energy propagation of waves in a dispersive medium may be identified as the ratio of the energy flux to the energy density, which is the group velocity, written here as $\beta_{\rm g}c$. It is shown in Appendix~\ref{sect:app} that  $\beta_{\rm g}$ for the L~mode is given by
\be
\beta_{\rm g}=\frac{d[z^2W(z)]/dz}{z\,dW(z)/dz}=z\left[1-2R_{\rm L}(z)\right].
\label{betag1}
\ee 

\begin{figure}
\begin{center}
\psfragfig[width=1.0\columnwidth]{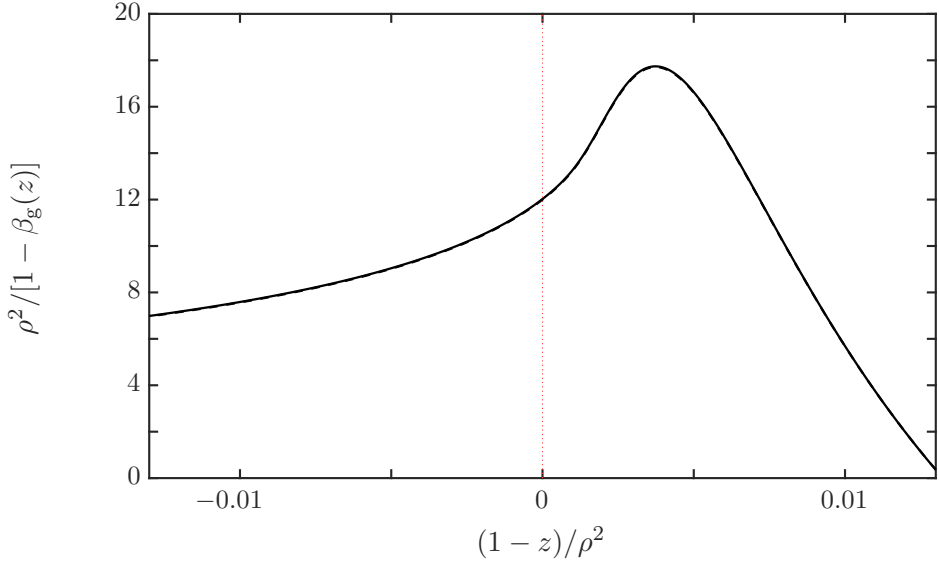}
 \caption{The function $\rho^2/[1-\beta_{\rm g}(z)]$ is plotted over $ (1-z)/\rho^2 $ for $ \rho = 0.1 $ (solid) and 0.01 (dashed). The two curves are indistinguishable over the scale shown. The thin dotted vertical line indicates the light line.}
 \label{fig:betag}  
 \end{center}
\end{figure}

Approximations for $z\gg1$ and $z=1$ are
\be
\beta_{\rm g}(z)\approx \frac{1}{z}\frac{3z^2 + 1}{z^2+3},
\qquad
\beta_{\rm g}(1)\approx1-\frac{1}{12\langle\gamma\rangle^2},
\label{betag3}
\ee  
respectively. Thus, the group speed is zero at the cutoff frequency $ \omega_x $ (where $z\to\infty$), and it increases with decreasing $z>1$, approaching unity, but remaining subluminal ($\beta_{\rm g}<1$) for $z\to1$. For $1>z\gg z_{\rm m}$ the final expression in equation (\ref{betag1}), and the form of $R_{\rm L}(z)$ shown in Figure~\ref{fig:1/RL}, imply that $\beta_{\rm g}$ gets closer to unity with decreasing $z$, reaching its maximum value at the maximum of $R_{\rm L}(z)$; it then decreases to zero at $z=z_{\rm m}$, as shown in Figure~\ref{fig:betag}. It follows from the scaling in the figure that $(1-\beta_{\rm g})/(1-z)$ is approximately independent of $\rho$ for $\rho\ll1$.

We do not discuss the region $z<z_{\rm m}$ in detail, but remark that equation (\ref{betag1}) has properties that preclude the interpretation of $\beta_{\rm g}$ as the speed of (wave) energy propagation for $z<z_{\rm m}$: it becomes negative for $z<z_{\rm m}$, singular at the zero of $dW(z)/dz$, where it changes sign to very large (superluminal group speed) and positive for smaller $z$.

\subsection{Wave properties for $1\geq z\gg z_{\rm m}$}

The only waves that can satisfy the resonance condition for a beam-driven instability in a pulsar plasma have properties that are quite different from those of Langmuir waves that can be beam-driven in a nonrelativistic plasma. Here we comment briefly on three unusual properties of the parallel propagating waves in the L~mode: the dispersion relation, the ratio of electric to total energy and the group speed.

In is convenient to write the condition $z=\beta_{\rm b}$ for resonance with a beam with speed $\beta_{\rm b}$ in terms of the corresponding Lorentz factor $\gamma_\phi=\gamma_{\rm b}$. For beam-driven growth to overcome Landau damping in the background plasma requires $\gamma_\phi\gg\gamma_{\rm m}$, with $\gamma_{\rm m}\approx6\langle\gamma\rangle$ for $\rho\lesssim1$. In the example shown in Figure~\ref{fig:log1-z}, this corresponds to a tiny range of the solid black dispersion curve: to the right of $(1-z)/\rho^2 = 0 $ and to the left of the peak in the curve at $z=z_{\rm m}$. Over this range $\omega_{\rm L}(z)$ is a rapidly decreasing function of $z$, with
\be
\frac{1}{\omega_{\rm L}(z)}\frac{d\omega_{\rm L}(z)}{dz}=\frac{1}{2z^2W(z)}\frac{d[z^2W(z)]}{dz}=\frac{1}{z}\left[1-\frac{1}{2R_{\rm L}(z)}\right]\approx-\frac{1}{2R_{\rm L}(z)},
\label{Ddr}
\ee
where the approximation applies for $\gamma_\phi\gg\gamma_{\rm m}$. With $1/2R_{\rm L}(z)$ ranging between $12\langle\gamma\rangle^2$ and $19\langle\gamma\rangle^2$ for $z<1$, cf. Figure~\ref{fig:1/RL}, it follows that the frequency $\omega_{\rm L}(z)$ is a very strong function of $z$ over the tiny range of $z$ where resonant wave growth is possible. This is quite different from Langmuir waves, for which the frequency is only a weak function over a wide range of phase speeds. It is also quite different from ion (or electron) acoustic waves, for which the phase speed is approximately equal to the ion (or electron) sound speed.

A second feature that is not ``Langmuir-like'' is that the ratio of electric to total energy $R_{\rm L}(z)$ is very small for L~mode waves in the range $1>z\gg z_{\rm m}$. In contrast, for Langmuir waves the ratio is approximately $1/2$, corresponding to approximate equipartition between electric energy and kinetic energy of forced motions in the wave. The very small value of $R_{\rm L}(z)$ is unusual when compared with Langmuir waves, but not when compared with ion acoustic waves, which have $R_{\rm L}\approx k^2\lambda_{\rm De}^2\ll1$, where $k$ is the wavenumber and $\lambda_{\rm De}$ is the electron Debye length. 

The group speed for L~waves is only marginally subluminal, $1-\beta_{\rm g}(z)\ll1$, in the relevant range, $1>z\gg z_{\rm m}$. The small group speed for Langmuir waves implies that the wave energy that grows in a given region due to a beam-driven instability remains localized to that region as the beam propagates through it. However, $\beta_{\rm g}(z)\approx1$ for the relevant L~mode waves implies that the wave energy propagates away at nearly the speed of light, impeding any wave growth. 

\section{Discussion and Conclusions}
\label{sect:conclusions}

We discuss wave dispersion in a pulsar plasma, defined as a strongly magnetized pair plasma with a 1D J\"uttner distribution. We find that the relativistic plasma dispersion function (RPDF) scales in a characteristic way with inverse temperature in the highly relativistic case, $\rho\ll1$. The favored case $\rho\approx1$ is more closely analogous to the highly relativistic case than to the nonrelativistic case, $\rho\gg1$.

Wave dispersion in the rest frame of a pulsar plasma is strongly modified compared with wave dispersion in more familiar nonrelativistic plasmas. One difference is due to the Alfv\'en speed, $\beta_{\rm A}c$, being highly relativistic $\beta_{\rm A}\gg1$. It is well known that the displacement current then plays an important role, and that the phase speed of Alfv\'en and magnetoacoustic waves (in a cold plasma) is determined by $\beta_0=\beta_{\rm A}/(1+\beta_{\rm A}^2)^{1/2}<1$ rather than $\beta_{\rm A}$. The parameter $z_{\rm A}$ used here is approximately equal to $\beta_0$ in a highly relativistic plasma, cf. (\ref{Lambdaij1}). Another difference is that the relativistic plasma dispersion function (RPDF) becomes extremely sharply peaked, compared with its nonrelativistic counterpart, and this has a large effect on some aspects of the wave dispersion. A third difference concerns the cross-over between the dispersion curves for the L~and A~mode. A cross-over necessarily occurs in the cold plasma case, and it leads to a resonance ($N^2\to\infty$, $z\to0$) in the Alfv\'en mode. In the relativistic case, the cross-over occurs only for $z_{\rm A}>z_0$ or $\beta_{\rm A}>\gamma_0\gg1$, which condition is satisfied for plausible pulsar parameters. Then there is a turnover, rather than a resonance, in the Alfv\'en mode. We note that for $z_{\rm A}<z_0$ there is no cross-over for $\theta=0$; in this case the modes reconnection for $b\tan^2\theta=z_0^2-z_{\rm A}^2$, forming (the O~mode and) an intrinsically oblique Alfv\'en-like mode (which we do not discuss in this paper). 

A motivation for the investigation reported here was an argument that beam-driven RPE is the most plausible emission mechanism for the nanoshots from the Crab pulsar \citep{EH16}. This suggestion is based on the assumption that the beam causes Langmuir-like waves to grow to a very high level in a localized region, with some nonlinear plasma process partly converting the wave energy into escaping radio emission \citep{W97,W98}. It is implicit in this model that Langmuir-like waves exist in the plasma, and that they are relatively slowly propagating, such that their group speed is small compared with the beam speed.  Similar assumptions were made in most early models of RPE, where Langmuir-like waves were simply assumed to exist. However, none of the wave modes discussed in the present paper satisfy all the requirements for ``Langmuir-like'' waves implicit in these models, specifically waves that can grow due to resonance with a beam and that are slowly propagating compared with the beam. Such resonance requires $\gamma_\phi=\gamma_{\rm b}$, where $\gamma_\phi$ and $\gamma_{\rm b}$ are the Lorentz factors corresponding to the phase speed of the  wave and the beam speed, respectively. The only relevant subluminal waves are the O~mode at sufficiently small angles $\theta$, which has $\gamma_\phi>\beta_{\rm A}$, and the Alfv\'en mode which has $\gamma_\phi\approx\beta_{\rm A}$ at low frequencies and turns over at higher frequencies, where $\gamma_\phi$ decreases from $\approx\beta_{\rm A}$ to order of $\langle\gamma\rangle$, as shown in Figure~\ref{fig:log1-z}. The nearest approximation to a ``Langmuir-like'' mode is the Alfv\'en mode near the turnover at $z=z_{\rm m}$, where its group speed is zero. Although beam-driven growth of Alfv\'en waves has been discussed as a possible pulsar emission mechanism  \citep{TK72,Letal82,MG99,L00}, the role that this turnover might play has not been discussed.

A discussion of the implications of these results for pulsar radio emission requires detailed estimates of the plasma parameters, which we propose to give elsewhere. However, simple estimates suggest that RPE encounters serious difficulties.  The conditions $\gamma_{\rm b}=\gamma_\phi$ and $\gamma_\phi>\beta_{\rm A}$ can be satisfied only for an extremely high-energy beam, due to $\beta_{\rm A}$ being extremely large in a pulsar magnetosphere. For example, consider an estimate of $\beta_{\rm A}^2=\Omega_{\rm e}^2/\omega_{\rm p}^2\langle\gamma\rangle$ with  $\Omega_{\rm e}$ evaluated for a magnetic field $B=10^8\,$T and the plasma frequency given by $\omega^2_{\rm p}\approx\kappa\Omega_{\rm e}\Omega_*/\gamma_{\rm s}$, with $\kappa$ the multiplicity (ratio of number density to the corotation charge density$/e$), $\Omega_*=2\pi/P$ the rotation frequency of the star and $\gamma_{\rm s}$ the Lorentz factor of the transformation between the rest frame and the pulsar frame. This gives $\beta_{\rm A}^2$ of order $10^{19}\beta_{\rm s}P/\kappa\langle\gamma\rangle$ near the surface of the star. For values $P=1\,$s, $\kappa=10^5$, $\gamma_{\rm s}=10^3$, $\langle\gamma\rangle=10$ this implies $\beta_{\rm A}^2$ of order $10^{15}$. For a dipolar magnetic field $B\propto1/r^3$ one has $\beta_{\rm A}^2\propto1/r^3$, but $\beta_{\rm A}$ remains very large, except perhaps near the light-cylinder radius, $r_{\rm L}=Pc/2\pi$. The most energetic beams considered plausible have Lorentz factors of order $10^6$--$10^7$ in the pulsar frame, and hence of order $10^3$--$10^4$ in the rest frame. We conclude that for the O~mode, the resonance condition fails to be satisfied by many orders of magnitude for a source near the star. Moreover, O~mode waves are not Langmuir-like in that their group speed is not slow but is close to the speed of light, and as they propagate the angle $\theta$ increases due to the curvature of the field lines, so that $\gamma_\phi$ increases and the waves quickly move out of resonance and become superluminal. We conclude that the assumption that the energy in these waves remains localized and builds up due to the beam propagating through the location is not justified. 

It is possible in principle for waves in the Alfv\'en mode to grow. However, for Alfv\'en waves at low frequencies, which satisfy the dispersion relation $z\approx z_{\rm A}$, the resonance condition is $\gamma_{\rm b}\approx\beta_{\rm A}$ and the foregoing discussion of the O~mode also applies to such Alfv\'en waves: the resonance condition cannot be satisfied under plausible conditions anywhere in the pulsar magnetosphere. At higher frequencies, as shown in Figure~\ref{fig:log1-z}, the dispersion relation turns over, with the turnover frequency decreasing with increasing $\theta$; with $(1-z)/\rho^2\approx\langle\gamma\rangle^2/2\gamma_\phi^2$, the turnover occurs at $\gamma_\phi\approx6\,\langle\gamma\rangle$, and Landau damping becomes important at about this and smaller $\gamma_\phi$. It follows that near the turnover, a beam with $\gamma_{\rm b}\approx6\,\langle\gamma\rangle$ can cause Alfv\'en waves to grow. Moreover, the group speed is small near the turnover, where $\partial\omega/\partial z$ passes through zero, favoring energy in these waves building up rather than propagating away. The properties of Alfv\'en waves near the turnover frequency were not taken into account in earlier models for beam-driven growth of Alfv\'en waves in a pulsar plasma \citep{TK72,Letal82,MG99,L00}. 

The discussion of the wave dispersion in this paper applies to waves in the rest frame of a pulsar plasma at frequencies well below the electron cyclotron frequency. Our choice of the rest frame of the plasma is convenient for formal purposes, but the relevant frame from an observational viewpoint is the pulsar frame, in which the pair plasma is streaming outward with Lorentz factor $\gamma_{\rm s}$. In Paper~2 we discuss the Lorentz transformation to the pulsar frame, and consider various aspects of the wave dispersion in that frame and in other frames where there are relative streaming motions between different distributions of particles.

\section*{Acknowledgments} 
We thank Mike Wheatland, referee V. S. Beskin and an anonymous referee for helpful comments on the manuscript. The research reported in this paper was supported by the Australian Research Council through grant DP160102932.

\bibliographystyle{jpp}

\bibliography{Pulsar_radio_Refs}

\appendix

\section{Dispersion equation}
\label{app:dispersion}

Here we present the derivation of nonzero components of $ \Lambda_{ij}(\omega, {\bi k}) $, defined in~\eqref{eq:Lambdaij}, and discuss our assumptions.

\subsection{Dielectric tensor}
Consider a plasma that is composed of electrons, $\epsilon=-1$, and positrons, $\epsilon=+1$. The general form for the dielectric tensor derived using kinetic theory involves a sum over $\epsilon=\pm1$. For a 1D distribution it is convenient to replace the conventional distribution function, $f_{\epsilon}(p_\perp,p_\parallel)$, and the integral over $2\pi dp_\parallel dp_\perp\,p_\perp$ by the 1D distribution $g_{\epsilon}(u)$ with $p_\perp=0$, $p_\parallel=mcu$, $u=\gamma\beta$, and the integral over $du$. The dielectric tensor is given by \citep{M08,M13}
\be
K_{ij}(\omega,{\bi k})=\delta_{ij}+\frac{\Pi_{ij}(\omega,{\bi k})}{\omega^2},
\qquad
\Pi_{ij}(\omega,{\bi k})=-\sum_{\epsilon}\omega_{p\epsilon}^2
\left\langle \frac{A_{ij}(\omega,{\bi k};\beta)}{\gamma}\right\rangle_{\epsilon},
\label{NPP1}
\ee
with $A_{ij}(\omega,{\bi k};\beta)\to A_{ij}$ given by equation (\ref{Aij1}) below, and with plasma frequency $\omega_{p\epsilon}^2=e^2n_{\epsilon}/\varepsilon_0m$, where $n_{\epsilon}$ is the number density of electrons or positrons in the plasma rest frame. The average $ \av{Q} $ of any function $Q$ of $u$ is written as
\be
    n_{\epsilon}\langle Q\rangle_{\epsilon}
        = \int{\rm d}u\,Q\,g_{\epsilon}(u),
    \label{average}
\ee 
which defines the number density, $n_\epsilon$, for $ Q = 1 $.

With the magnetic field along the 3-axis and the wave vector in the 1-3 plane, we introduce the notation
\be
    {\bi k}=(k_\perp,0,k_\parallel )=\frac{\omega}{zc}(\tan\theta,0,1),
    \quad{\rm with}\quad
    z = \frac{\omega}{k_\parallel c}.
    \label{bik}
\ee
The components of the tensor $ A_{ij} $ in equation (\ref{NPP1}) are
\begin{equation}
\begin{gathered}
    A_{11} = \frac{\omega_0^2}{\omega_0^2-\Omega^2},\quad
    A_{12} = \rmi\epsilon\frac{\omega_0\Omega}{\omega_0^2-\Omega^2},\quad
    A_{23} = -\rmi\epsilon\frac{\omega\Omega}{\omega_0^2-\Omega^2}\frac{\beta\tan\theta}{z}\\
    A_{33} = \frac{\omega^2}{\gamma^2\omega_0^2} +\frac{\omega^2}{\omega_0^2-\Omega^2}\left(\frac{\beta\tan\theta}{z}\right)^2,\quad
    A_{13} = \frac{\omega_0\omega}{\omega_0^2-\Omega^2}\frac{\beta\tan\theta}{z},
\end{gathered}
\label{Aij1}
\end{equation}
with $ A_{22} = A_{11} $, $ A_{31} = A_{13} $, $ A_{21} = -A_{12} $, $ A_{32} = -A_{23} $, $\omega_0=\omega-k_\parallel v_\parallel=\omega(z-\beta)/z$ and $\Omega=\Omega_{\rm e}/\gamma$, where $\Omega_{\rm e}=eB/m$ is the electron cyclotron frequency.

\subsection{Low-frequency and non-gyrotropic approximations}

The expression (\ref{NPP1}) with (\ref{Aij1}) is exact for a 1D distribution, and simplifying assumptions and approximations need to be made in applying it to a pulsar plasma. One simplifying assumption is that there is a single distribution of pairs; this assumption, which is made here, needs to be relaxed to discuss  any beam-driven instability. Two other simplifying approximations are made here: the low-frequency limit and the non-gyrotropic approximation.

\subsubsection{Low-frequency limit}
\label{app:low_frequency}

Pulsar radio emission is thought to be generated in regions where the wave frequency is much smaller than the cyclotron frequency. We assume the low-frequency approximation in the form $\omega_0\ll\Omega$. On expanding in $\omega_0/\Omega$, the leading terms in the non-gyrotropic components are, for a 1D distribution,
\begin{equation}
\begin{gathered}
    A_{11} = A_{22} = - \frac{\gamma^2\omega^2}{\Omega_{\rm e}^2}\left(\frac{z-\beta}{z}\right)^2,\quad
    A_{13} = A_{31} = -\frac{\gamma^2\omega^2}{\Omega_{\rm e}^2}\frac{z-\beta}{z}\frac{\beta\tan\theta}{z},\\
    A_{33} = \left(\frac{z}{\gamma(z-\beta)}\right)^2-\frac{\gamma^2\omega^2}{\Omega_{\rm e}^2}\left(\frac{\beta\tan\theta}{z}\right)^2.
\end{gathered}
\label{Aij2}
\end{equation}
On averaging over the distribution function, the term $\propto1/(z-\beta)^2$ leads to the RPDF discussed below, and the other terms involve $\langle\gamma\rangle,\langle\gamma\beta\rangle,\langle\gamma\beta^2\rangle$. A simplifying assumption in the rest frame is that $g(-u)=g(u)$ is an even function, implying $\langle\gamma\beta\rangle=0$. One has $\langle\gamma\beta^2\rangle=\langle\gamma\rangle - \av{1/\gamma} \approx \av{\gamma} $ for $\langle\gamma\rangle\gg1$. It follows that for $\langle\gamma\rangle\gg1$, apart from the RPDF, the only important average is $\langle\gamma\rangle$. 

\subsubsection{Non-gyrotropic approximation}
\label{app:non_gyrotropic}

The gyrotropic terms, $A_{12}=-A_{21}$, $A_{32}=-A_{23}$, in the low-frequency approximation are given by
\be
A_{12}=-\rmi\epsilon\frac{\gamma\omega}{\Omega_{\rm e}}\frac{z-\beta}{z},
\qquad
A_{23}=\rmi\epsilon
\frac{\gamma\omega}{\Omega_{\rm e}}\frac{\beta\tan\theta}{z}.
\label{gyrotropic}
\ee
After substitution into equation (\ref{NPP1}) these terms are summed over the electrons and positrons giving contributions proportional to $ (n_+ - n_-) $ and $ (n_+\langle\beta\rangle_+ - n_-\langle\beta\rangle_-) $, that is to the charge density and current density, respectively, both of which are nonzero in a pulsar plasma. However, these terms may be regarded as of first order, in comparison with the terms in equation (\ref{Aij2}), in an expansion in $1/\kappa$, where $\kappa=(n_+ + n_-)/|n_+ - n_-|$ is the multiplicity, and then these terms contribute to the dispersion equation only to second order in $1/\kappa$. The non-gyrotropic approximation corresponds to neglecting the gyrotropic terms. This is equivalent to assuming that the distribution functions for the electrons and positrons are identical, $g_{+}(u)=g_{-}(u)$. The subscript $\epsilon$ is redundant in the non-gyrotropic approximation, and is omitted in the following discussion. It is essential to relax the non-gyrotropic assumption in order to discuss the ellipticity of the polarization of the natural modes \citep[e.g.,][]{LMF02,LM04a,BP12}, but we do not do so here.

\subsubsection{Dielectric tensor for pulsar plasma}

With these assumptions, the non-gyrotropic components of the dielectric tensor in the rest frame of a pulsar plasma reduce to, $ K_{ij}(\omega, {\bi k}) \to K_{ij} $, \citep{MGKF99}
\begin{equation}\label{eq:Kij0}
\begin{gathered}
    K_{11} = K_{22} = 1 + \frac{\omega_{\rm p}^2}{\Omega_e^2}\frac{1}{z^2}\left\langle\gamma(z - \beta)^2\right\rangle,\quad
    K_{13} = K_{31} = \frac{\omega_{\rm p}^2}{\Omega_e^2}\frac{\tan\theta}{z^2}\left\langle\gamma\beta(z - \beta)\right\rangle,\\
    K_{33} = 1 - \frac{\omega_{\rm p}^2}{\omega^2}z^2 W(z) + \frac{\omega_{\rm p}^2}{\Omega_e^2}\frac{\tan^2\theta}{z^2}\left\langle\gamma\beta^2\right\rangle.
\end{gathered}
\end{equation}
The RPDF $ W(z) $ is defined by
\be
W(z) =\frac{1}{n}\int{\rm d} u\frac{1}{\beta - z}\frac{\rmd g(u)}{\rmd u}.
\label{Wz}
\ee
In the rest frame of the plasma $ g(u) $ is an even function of $ \beta $ which then implies that~\eqref{eq:Kij0} may be expressed as
\begin{equation}
\begin{gathered}
    K_{11} = K_{22} = 1+\frac{1}{\beta_{\rm A}^2}\left(1+\frac{\Delta\beta^2}{z^2}\right),\quad
    K_{33} = 1 - \frac{\omega_{\rm p}^2}{\omega^2}z^2W(z)+\frac{\Delta\beta^2\tan^2\theta}{\beta_{\rm A}^2z^2},\\
    K_{13} = K_{31} = -\frac{\Delta\beta^2\tan\theta}{\beta_{\rm A}^2z^2},\quad
    \beta_{\rm A}^2 = \frac{\Omega_{\rm e}^2}{\omega_{\rm p}^2\langle\gamma\rangle},\quad
    \Delta\beta^2 = \frac{\langle\gamma\beta^2\rangle}{\langle\gamma\rangle}=1-\frac{\langle\gamma^{-1}\rangle}{\langle\gamma\rangle},
\end{gathered}
\label{Kij1}
\end{equation}
where we use $ \av{Q} = 0 $ for $ Q $ any odd function of $ \beta $.

\subsection{Dispersion equation for a pulsar plasma}

The dispersion equation for waves in a plasma is given by setting the determinant of the matrix form of $\Lambda_{ij}$ to zero. The nonzero components of $\Lambda_{ij}$ are obtained using~\eqref{we1} and~\eqref{eq:Kij0} as
\begin{equation}\label{eq:Lambdaij}
\begin{gathered}
    \Lambda_{11} 
        = 1 - \frac{1}{z^2} + \frac{\omega_{\rm p}^2}{\Omega_e^2}\frac{1}{z^2}\left\langle\gamma(z - \beta)^2\right\rangle,\quad
    \Lambda_{13} 
        = \frac{\tan\theta}{z^2}\left[1 + \frac{\omega_{\rm p}^2}{\Omega_e^2}\left\langle\gamma\beta(z - \beta)\right\rangle\right],\\
    \Lambda_{22}
        = \Lambda_{11} - \frac{\tan^2\theta}{z^2},\quad
    \Lambda_{33} 
        = 1 - \frac{\omega_{\rm p}^2}{\omega^2}z^2W(z) - \frac{\tan^2\theta}{z^2}\left[1 - \frac{\omega_{\rm p}^2}{\Omega_e^2}\left\langle\gamma\beta^2\right\rangle\right],
\end{gathered}
\end{equation}
with $ \Lambda_{31} = \Lambda_{13} $. Noting that $ \langle Q\rangle = 0 $ if $ Q $ is an odd function of $ \beta $ allows us to write~\eqref{eq:Lambdaij} as
\begin{equation}
\begin{gathered}
    \Lambda_{11}=a-\frac{b}{z^2}, \quad
    \Lambda_{22}=\Lambda_{11}-\frac{\tan^2\theta}{z^2}, \quad
    \Lambda_{33}=1-\frac{\omega_{\rm p}^2}{\omega^2}z^2W(z)-\frac{b\tan^2\theta}{z^2},\\
    \Lambda_{13}=\frac{b\tan\theta}{z^2},\quad
    a=1+\frac{1}{\beta_{\rm A}^2}, \quad
    b=1-\frac{\Delta\beta^2}{\beta_{\rm A}^2},\quad
    z_{\rm A}^2=\frac{b}{a}=\frac{\beta_{\rm A}^2-\Delta\beta^2}{1+\beta_{\rm A}^2}.
    \label{Lambdaij1}
\end{gathered}
\end{equation}
The parameter $z_{\rm A}$ may be interpreted as the relativistic Alfv\'en speed. The Lorentz factor corresponding to this speed is $\gamma_{\rm A}$, given by
\be
\gamma_{\rm A}^2=\frac{1}{1-z_{\rm A}^2}
=\frac{1+\beta_{\rm A}^2}{1+\Delta\beta^2},
\label{gammaA}
\ee
with $\gamma_{\rm A}\approx\beta_{\rm A}$ in the highly relativistic limit. Equation~(\ref{Lambdaij1}) reproduce expressions given by \citet{MGKF99}, except for the correction of an error in $\Lambda_{22}$ ($-\sin^2\theta$ is replaced by $+\tan^2\theta$). 

In the cold plasma limit, $ \rho \to \infty $ and $ \langle\gamma\rangle \to 1 $, and infinite magnetic field, $ \Omega_e \to \infty $ and $ \beta_A \to \infty$, the expressions in~\eqref{Lambdaij1} may be approximated as
\begin{equation}
\begin{gathered}
    \Lambda_{11}=a-\frac{1}{z^2}, \quad
    \Lambda_{22}=\Lambda_{11}-\frac{\tan^2\theta}{z^2}, \quad
    \Lambda_{33}=1-\frac{\omega_{\rm p}^2}{\omega^2}-\frac{\tan^2\theta}{z^2},\\
    \Lambda_{13}=\frac{\tan\theta}{z^2},\quad
    a=1, \quad
    b=1,\quad
    z_{\rm A}^2=1,
    \label{Lambdaij1b}
\end{gathered}
\end{equation}
where we use $ z^2W(z) \to 1 $ in the cold plasma limit.

\section{RPDF for $ |z| \leq 1 $}
\label{sect:appRPDF}
We may write the RPDF~\eqref{Wz} as
\begin{align}
W(z) & = \lim_{\delta \to 0}\left(\int_{-1}^{z-\delta} + \int_{z - \delta}^{z + \delta} + \int_{z + \delta}^{1} \right){\rm d} \beta\, \frac{1}{\beta-z}\frac{\rmd g(u)}{\rmd \beta}\\
	& = i\pi \left.\frac{k_\parallel}{|k_\parallel|}\frac{\rmd g(u)}{\rmd \beta}\right|_{\beta = z} + \lim_{\delta \to 0}\left(\int_{-1}^{z-\delta} + \int_{z + \delta}^{1} \right)\rmd \beta\, \frac{1}{\beta-z}\frac{{\rm d} g(u)}{\rmd \beta},
\end{align}
where, as per the Landau prescription, we integrate below the singularity at $ \beta = z $ in the complex $ \beta $ plane. The integral over $ z - \delta \leq \beta \leq z + \delta $ is performed in a positive sense and contributes $ i\pi $ times the residue. The remaining integrals may be partially integrated to give
\begin{equation}
\lim_{\delta \to 0}\left(\int_{-1}^{z-\delta} + \int_{z + \delta}^{1} \right)\rmd \beta\, \frac{1}{\beta-z}\frac{{\rm d} g(u)}{\rmd \beta}
= \lim_{\delta \to 0}\left\{-2\frac{\left.g(u)\right|_{\beta = z}}{\delta} + \left(\int_{-1}^{z-\delta} + \int_{z + \delta}^{1} \right){\rm d} \beta\, \frac{g(u)}{(\beta-z)^2}\right\},
\end{equation}
where we use $ \left.g(u)\right|_{\beta = z\pm \delta} \approx \left.g(u)\right|_{\beta = z} $ for sufficiently small $ \delta $. Noting that
\begin{equation}
\wp \int_{-1}^{1}{\rm d}\beta\, \frac{1}{(\beta - z)^2} = -2\gamma_\phi^2 + \lim_{\delta \to 0}\frac{2}{\delta},
\end{equation}
then gives~\eqref{Wz2}.
\section{Group velocity}
\label{sect:app}

Let the dispersion equation, $K_{33}=0$, for the L~mode be written $K=0$ with
\be
K=1-\frac{\omega_{\rm L}^2(z)}{\omega^2},
\qquad
\omega_{\rm L}^2(z)=\omega_{\rm p}^2z^2W(z).
\label{app1}
\ee
The chain rule implies
\be
\left.\frac{\partial\omega}{\partial k_\parallel}\right|_{K}
\left.\frac{\partial k_\parallel}{\partial K}\right|_{\omega}
\left.\frac{\partial K}{\partial \omega}\right|_{k_\parallel}=-1,
\qquad
\left.\frac{\partial\omega}{\partial k_\parallel}\right|_{K}=-
\left.\left.\frac{\partial K}{\partial k_\parallel}\right|_{\omega}\right/
\left.\frac{\partial K}{\partial \omega}\right|_{k_\parallel},
\label{app2}
\ee
to be evaluated at $K=0$. Using $z=\omega/k_\parallel c$ and equation (\ref{app1}) one finds
\be
\left.\frac{\partial K}{\partial k_\parallel}\right|_{\omega}=
\frac{\omega_{\rm p}^2}{\omega^2}\frac{cz^2}{\omega}\frac{d[z^2W(z)]}{dz},
\qquad
\left.\frac{\partial K}{\partial \omega}\right|_{k_\parallel}=-\frac{\omega_{\rm p}^2}{\omega^2}\frac{z^3}{\omega}
\frac{dW(z)}{dz}.
\label{app3}
\ee
The ratio of electric to total energy becomes
\be
R_{\rm L}=\left[\omega\left.\frac{\partial K}{\partial \omega}\right|_{k_\parallel}\right]^{-1}=-\frac{\omega^2}{\omega_{\rm p}^2z^3dW(z)/dz}=
-\frac{W(z)}{zdW(z)/dz}.
\label{app4}
\ee
The group speed becomes
\be
\beta_{\rm g}=\frac{d[z^2W(z)]/dz}{z\,dW(z)/dz} = z[1 - 2 R_{\rm L}(z)].
\label{app5}
\ee

\section{Approximation of $ z^2 W(z) $}
\label{app:C}

For $ z \gg 1 $ we may write
\begin{equation}
    \frac{z^2}{\gamma^3(\beta - z)^2}
        = \frac{1}{\gamma^3}\sum_{s = 0}^{\infty}(s+1)(\beta/z)^s,
\end{equation}
which implies, after swapping the order of intergation and summation,
\begin{equation}
    z^2W(z)  = \sum_{s = 0}^{\infty}(s+1)\frac{\aV{\beta^s/\gamma^3}}{z^s},
\end{equation}
with
\begin{equation}
   \aV{\beta^s/\gamma^3}
        =
        \begin{dcases}
        \sum_{j = 0}^{k}\frac{k!(-1)^j}{(k-j)!j!}\aV{\frac{1}{\gamma^{2j+3}}}, \quad & {\rm for} \quad s = 2k,\\
        0, \quad & {\rm for} \quad s = 2k + 1.
    \end{dcases}
\end{equation}
These then give the first case in~\eqref{Wza}. 

Expansion about $ z = 1 $ gives
\begin{align}
    \frac{z^2}{\gamma^3(\beta - z)^2}
        & = \gamma(1 + \beta)^2\left[1 + \sum_{s = 1}^\infty \gamma^{2s}(1-z)^s (2\beta + (s-1)\beta^2)(1 + \beta)^s\right]\\
        & = \gamma(1 + \beta)^2\left[1 + \sum_{s = 1}^\infty \gamma^{2s}(1-z)^s (2\beta + (s-1)\beta^2) \sum_{k = 0}^s \frac{s! \beta^k}{(s-k)!k!}\right],
\end{align}
which implies
\begin{equation}\label{eq:z2Wa}
    z^2W(z)
        = 2\aV{\gamma} - \aV{1/\gamma} + \sum_{s = 1}^\infty \sum_{k = 0}^s \frac{s! (1-z)^s}{(s-k)!k!}\aV{\gamma^{2s+1}(2\beta + (s-1)\beta^2)(1+\beta)^2\beta^k}.
\end{equation}
The average quantity may be expressed as
\begin{equation}
    \sum_{r = 0}^{j + 1}\frac{(-1)^r(j+1)!}{(j + 1 - r)!r!}
    \begin{dcases}
        2(s + 1)\aV{\gamma^{2(s-r)+1}} - (s-1)\aV{\gamma^{2(s-r)-1}}, \quad & {\rm for} \quad k = 2j,\\
        2(s + 1)\aV{\gamma^{2(s-r)+1}} - 2s\aV{\gamma^{2(s-r)-1}}, \quad & {\rm for} \quad k = 2j + 1.
    \end{dcases}
\end{equation}
Each term inside the sum in~\eqref{eq:z2Wa} is then, to highest order in $ \av{\gamma} \gg 1 $, equals to
\begin{equation}
    \frac{s!(2s+2)!}{(s-k)!k!}(1-z)^s\av{\gamma}^{2s+1},
\end{equation}
where we use $ \av{\gamma^n} = n!\av{\gamma}^n $ for $ \av{\gamma} \gg 1 $. The series appears to have a radius of convergence of zero. This leads us to suggest that the approximation given in~\eqref{Wza} for the case $ |1 - z| \ll 1 $ is at best questionable and that it should not be used.

\section{List of notation}

\begin{table}
\centering
\caption{List of common symbols and parameters. The values each parameter takes may vary depending on the context.}
\begin{tabular}{ll}
     {\bf Symbol} & {\bf Description}  \\
    \cline{1-2}
    &\\
    $ \av{Q} $ & Average value of quantity $ Q $\\
    $ a $ & Parameter $ a = 1 + 1/\beta_{\rm A}^2 $\\
    $ b $ & Parameter $ b = 1 - \Delta\beta^2/\beta_{\rm A}^2 $\\
    $ B $ & Magnetic field strength\\
    $ \beta $ & Particle velocity\\
    $ \beta_{\rm A} $ & Alfv\'en speed: $ \beta_{\rm A}^2 = \Omega_{\rm e}^2/\omega_{\rm p}^2\av{\gamma} $\\
    $ \beta_{\rm g} $ & Group velocity\\
    $ \Delta\beta^2 $ & Parameter $ \Delta\beta^2 = \av{\gamma\beta^2}/\av{\gamma} = 1 - \av{1/\gamma}/\av{\gamma} $\\
    $ {\bi e} $, $ e_i $, $ e_j $ & Polarization vector $ {\bi e } = (e_1, e_2, e_3) $\\
    $ g(u) $ & Particle distribution function\\
    $ \gamma $  & Lorentz factor $ \gamma = (1-\beta)^{-1/2} $ where $ \beta c $ is the particle speed\\
    $ \gamma_{\rm A} $ & Lorentz factor evaluated at $ \beta = z_{\rm A} $\\
    $ \gamma_{\rm s} $ & Streaming Lorentz factor of the distribution\\
    $ \gamma_\phi $ & Lorentz factor evaluated at $ \beta = z $ for $ |z| \leq 1 $\\
    $ {\bi k} $, $ k_\perp$, $ k_\parallel $ & Wave vector $ {\bi k} = (k_\perp, 0, k_\parallel) $.\\
    $ K_{ij}(\omega, {\bi k}) $ & Dielectric tensor\\
    $ \kappa $ & Multiplicity\\
    $ \Lambda_{ij}(\omega, {\bi k}) $ & Wave equation tensor\\
    $ n $ & Number density in the plasma rest frame\\
    $ N $ & Refractive index $ N = 1/z\cos\theta $\\
    $ \omega $ & Wave frequency\\
    $ \omega_1 $ & Wave frequency where $ \omega_{\rm L}(z) $ crosses the light line $ z = 1 $\\
    $ \omega_{\rm co} $ & Crossover frequency: wave frequency where $ \omega_{\rm L}(z) $ crosses the A~mode at $ z = z_{\rm A} $\\
    $ \omega_{\rm L}(z) $ & Dispersion relation of the L~mode\\
    $ \omega_{\rm p} $ & Plasma frequency in the rest frame of plasma\\
    $ \omega_{\rm x} $ & Cutoff frequency: $ \omega_{\rm x} = \omega_{\rm L}(\infty) $\\
    $ \Omega_{\rm e} $ & Electron cyclotron frequency\\
    $ P $, $ \dot{P} $ & Period and period derivative of the pulsar, respectively\\
    $ r $, $ r_{\rm L} $ & Radial distance and light cylinder radius: $ r_{\rm L} = Pc/2\pi $\\
    $ R_{\rm L}(z) $ & Ratio of electric to total energy\\
    $ \rho $ & Inverse temperature in units of energy: $ \rho = mc^2/T $\\
    $ T $ & Plasma remperature in units of energy\\
    $ T(z, \rho) $ & Relativistic plasma dispersion function\\
    $ \theta $ & Wave propagation angle\\
    $ u = \gamma\beta $ & Particle 4-speed\\
    $ V $ & Thermal velocity with $ \rho = c^2/V^2 $\\
    $ W(z) $ & Relativistic plasma dispersion function\\
    $ z $ & Phase velocity: $ z = \omega/ck_\parallel $\\
    $ z_{\rm A} $ & Dispersion relation of the A~mode: $ z = z_{\rm A} $ with $ z_{\rm A}^2 = b/a $\\
    $ z_{\rm Imin} $, $ z_{\rm e1,2} $ & Value of $ z $ where $ z^2\Im W(z) $ is minimum, and where $ \left|z^2\Im W(z)\right| = \left|z^2\Re W(z)\right| $, respectively\\
    $ z_{\rm m} $, $ z_0 $, $ z_{\rm min} $ & Value of $ z $ where $ z^2 \Re W(z) $ is maximum, zero and minimum, respectively
    
\end{tabular}
\end{table}

\end{document}